\documentclass[11pt, draftcls, onecolumn]{IEEEtran}

\usepackage{amsmath,amssymb,epsfig,ifpdf,cite}
\usepackage{threeparttable}
\usepackage{subfigure,color}

\newcommand{\DetCdFigDir}{./}
\newcommand{\TexComDir}{./}

\newcommand{\Append}[1]{{Appendix~\ref{#1}}}

\newcommand{\Fig}[1]{{Fig.~\ref{#1}}}

\newcommand{\Section}[1]{{Section~\ref{#1}}}
\newcommand{\Table}[1]{{Table~\ref{#1}}}

\newcommand{\CN}{\mathcal{CN}}

\newcommand{\Cov}{\textrm{cov}}

\newcommand{\Ex}{\mbox{$\mathbb{E}$}}

\newcommand{\Var}{\textrm{var}}

\newcommand{\Ecal}{\mathcal{E}}

\newcommand{\Gcal}{\mathcal{G}}

\newcommand{\Ncal}{\mathcal{N}}

\newcommand{\xvec}{\mathbf{x}}

\newcommand{\zerovec}{\mathbf{0}}

\newcommand{\AlphamCdd}{{\alpha_m^{{}_{\textrm{cdd}}}}}

\newcommand{\BOne}{{b_1}}
\newcommand{\BTwo}{{b_2}}

\newcommand{\CohBwCdd}{{B_{{}^{\textrm{C}}}^{{}^{\textrm{cdd}}}}}

\newcommand{\ExMaxCb}{{\Ex[\max_b C_b]}}

\newcommand{\GammanCdd}{{\gamma_n^{{}_{\textrm{cdd}}}}}
\newcommand{\GammanOneCdd}{{\gamma_\NOne^{{}_{\textrm{cdd}}}}}
\newcommand{\GammanTwoCdd}{{\gamma_\NTwo^{{}_{\textrm{cdd}}}}}

\newcommand{\gammaBar}{{\overline{\gamma}}}

\newcommand{\hCdd}{{h^{{}^{\textrm{cdd}}}}}

\newcommand{\him}{{h_{i,m}}}
\newcommand{\Hin}{{H_{i,n}}}
\newcommand{\HinOne}{{H_{i,\NOne}}}
\newcommand{\HinTwo}{{H_{i,\NTwo}}}

\newcommand{\HnCdd}{{H_n^{{}^{\textrm{cdd}}}}}
\newcommand{\HnOneCdd}{{H_{\NOne}^{{}^{\textrm{cdd}}}}}
\newcommand{\HnTwoCdd}{{H_{\NTwo}^{{}^{\textrm{cdd}}}}}

\newcommand{\NFb}{{N_{\hspace{-0.05cm}{}_{\textrm{FB}}}}}

\newcommand{\NOne}{{n_1}}
\newcommand{\Nrb}{{N_{\hspace{-0.05cm}{}_{\textrm{RB}}}}}

\newcommand{\NrbSq}{{N_{{}_{\textrm{RB}}}^2}}

\newcommand{\Nsc}{{N_{{}_{\textrm{SC}}}}}

\newcommand{\NTwo}{{n_2}}
\newcommand{\NTx}{{N_{{}_{\textrm{T}}}}}

\newcommand{\RhoRb}{{\rho_{{}_{\textrm{RB}}}}}
\newcommand{\RhoRbCdd}{{\rho_{{}_{\textrm{RB}}}^{{}_{\textrm{cdd}}}}}

\newcommand{\RhoSc}{{\rho_{{}_{\textrm{SC}}}}}
\newcommand{\RhoScCdd}{{\rho_{{}_{\textrm{SC}}}^{{}_{\textrm{cdd}}}}}

\newcommand{\RhoScSq}{{\rho_{{}_{\textrm{SC}}}^2}}

\newcommand{\RpfmKb}{{R^{\textrm{PFM}}_{k,b}}}
\newcommand{\RavgK}{{R^{\textrm{AVG}}_{k}}}
\newcommand{\Rsum}{{R_{{}_{\textrm{SUM}}}}}

\newcommand{\Srb}{{S_{{}_{\textrm{RB}}}}}
\newcommand{\SrbSq}{{S_{{}_{\textrm{RB}}}^2}}

\newcommand{\SumRb}{{\Psi_{\hspace{-0.5mm} {}_{\textrm{RB}}}}}
\newcommand{\SumRbCdd}{{\Psi_{\hspace{-0.5mm} {}_{\textrm{RB}}}^{{}^{\textrm{cdd}}}}}

\newcommand{\SumRbInAll}{{\sum_{b_1 = 1}^{\Nrb} \sum_{b_2 = 1}^{\Nrb}}}
\newcommand{\SumRbInv}{{\frac{1}{\SumRb(\Srb)}}}
\newcommand{\SumRbInvCdd}{{\frac{1}{\SumRbCdd(\Srb)}}}
\newcommand{\SumSc}{{\Psi_{\hspace{-0.5mm} {}_{\textrm{SC}}}}}
\newcommand{\SumScCdd}{{\Psi_{\hspace{-0.5mm} {}_{\textrm{SC}}}^{{}^{\textrm{cdd}}}}}

\newcommand{\SumScInTwoRb}{{\sum_{\substack{ n_1 = 1 + \\(b_1-1)\Srb }}^{b_1 \Srb} \sum_{\substack{ n_2 = 1 + \\(b_2-1)\Srb }}^{b_2 \Srb}}}
\newcommand{\SumScInv}{{\frac{1}{\SumSc(1,0,\Nsc)}}}

\newcommand{\SumScInvCdd}{{\frac{1}{\SumScCdd(1,0,\Nsc)}}}

\newcommand{\TauMaxi}{{\tau_{{}_{\textrm{max}},i}}}

\newcommand{\TauRms}{{\tau_{{}_{\textrm{rms}}}}}
\newcommand{\TauRmsSq}{{\tau_{{}_{\textrm{rms}}}^2}}
\newcommand{\TauRmsBar}{{\overline{\tau_{{}_{\textrm{rms}}}}}}
\newcommand{\TauRmsi}{{\tau_{{}_{\textrm{rms}},i}}}
\newcommand{\TauRmsiSq}{{\tau_{{}_{\textrm{rms}},i}^2}}
\newcommand{\TauRmsCdd}{{\tau_{{}_{\textrm{rms}}}^{{}_{\textrm{cdd}}}}}

\newcommand{\AWGN}{{\textrm{AWGN}}}

\newcommand{\BER}{{\textrm{BER}}}

\newcommand{\CC}{{\textrm{CC}}}

\newcommand{\CDD}{{\textrm{CDD}}}

\newcommand{\FFT}{{\textrm{FFT}}}

\newcommand{\MAC}{{\textrm{MAC}}}

\newcommand{\MMSE}{{\textrm{MMSE}}}

\newcommand{\MRC}{{\textrm{MRC}}}

\newcommand{\OFDM}{{\textrm{OFDM}}}
\newcommand{\OFDMA}{{\textrm{OFDMA}}}

\newcommand{\PDP}{{\textrm{PDP}}}
\newcommand{\PDF}{{\textrm{PDF}}}

\newcommand{\RB}{{\textrm{RB}}}

\newcommand{\RMS}{{\textrm{RMS}}}

\newcommand{\SC}{{\textrm{SC}}}

\newcommand{\SISO}{{\textrm{SISO}}}

\newcommand{\SNR}{{\textrm{SNR}}}

\newcommand{\eg}{{\it{e.g.,}}}
\newcommand{\Ei}{{\textrm{Ei}}}

\newcommand{\ie}{{\it{i.e.,}}}

\newcommand{\iid}{{\it{i.i.d.}}}

\newcommand{\OneColumn}[1]{}

\newcommand{\bItem}{\begin{itemize}}
\newcommand{\eItem}{\end{itemize}}

\newcommand{\bEq}{\begin{equation}}
\newcommand{\eEq}{\end{equation}}
\newcommand{\bEqn}{\begin{equation*}}
\newcommand{\eEqn}{\end{equation*}}
\newcommand{\bEqnarray}{\begin{eqnarray}}
\newcommand{\eEqnarray}{\end{eqnarray}}

\newcommand{\bCenter}{\begin{center}}
\newcommand{\eCenter}{\end{center}}

\newcommand{\SubPlus}{{\textrm{+}}}

\newcommand{\TauRmsj}{{\tau_{\mbox{\tiny rms},j}}}

\newcommand{\alphaim}{\alpha_{i,m}}

\newcommand{\alpham}{\alpha_{m}}

\newcommand{\DPerUser}{{D_{{}^\textrm{PerUser}}^{*}}}
\newcommand{\DPerUserVec}{{\underline{\mathbf{D}}_{{}^\textrm{PerUser}}^{*}}}
\newcommand{\DSumRate}{D_{{}^\textrm{SumRate}}^{*}}
\newcommand{\DSumRateVec}{{\underline{\mathbf{D}}_{{}^\textrm{SumRate}}^{*}}}
\newcommand{\DnumBc}{{D^{*}_{{}^{\textrm{B}_{{}_\textrm{c}}}}}}
\newcommand{\DnumMax}{{D^{*}_{{}^{\textrm{max}}}}}
\newcommand{\Dfix}{D^{\mbox{\tiny x}}}

\newcommand{\Dvec}{{\underline{\mathbf{D}}}}

\newcommand{\fcdd}{f_{\mbox{\tiny cdd}}}
\newcommand{\mucdd}{\mu_{\mbox{\tiny cdd}}}

\newcommand{\muone}{\mu^{\mbox{\tiny (1)}}}
\newcommand{\mutwo}{\mu^{\mbox{\tiny (2)}}}
\newcommand{\muw}{\mu_{\mbox{\tiny w}}}

        \newcommand{\TableColumnSizeOne}{10.0cm}
        \newcommand{\FigSizeNew}{10.0 cm}
        
        \newcommand{\FigWidth}{10.0 cm}
        \newcommand{\FigHeight}{6.25 cm}

\begin{document}

\title{On the optimal frequency selectivity to maximize multiuser diversity in an \OFDMA{} scheduling system}
\author
{
    Seong-Ho (Paul) Hur{\hspace{0.08cm}${}^*$}
    \IEEEmembership{Student Member, IEEE,}
    Bhaskar~D.~Rao,
    \IEEEmembership{Fellow, IEEE,}
    Min-Joong Rim,
    \IEEEmembership{Member, IEEE,}
    and~James~R.~Zeidler,
    ~\IEEEmembership{Fellow,~IEEE}

}
\maketitle

\begin{abstract}
\label{sAbstract}
We consider an orthogonal frequency division multiple access (\OFDMA) scheduling system.
A scheduling unit block consists of contiguous multiple subcarriers.
Users are scheduled based on their block average throughput in a proportional fair way.
The multiuser diversity gain increases with the degree and dynamic range of channel fluctuations.
However, a decrease of the block average throughput in a too much selective channel may lessen the sum rate as well.
In this paper, we first study optimal channel selectivity in view of maximizing the maximum of the block average throughput of an arbitrary user.
Based on this study, we then propose a method to determine a per-user optimal cyclic delay when cyclic delay diversity (\CDD) is used to enhance the sum rate by increasing channel selectivity for a limited fluctuating channel.
We show that the proposed technique achieves better performance than a conventional fixed cyclic delay scheme and that the throughput is very close to the optimal sum rate possible with \CDD{}.
\end{abstract}



\newpage

\section{Introduction}
\label{sIntroduction4FreqDiv}
Multiuser diversity is inherent in all multiuser wireless networks with independent fading among users \cite{bKnoppIcc95,bTseIsit97,bViswanathTit02}.
This diversity is exploited by scheduling the user with the best channel in a given time slot.
It leads to an increase of the system sum rate as the number of users increases \cite{bKnoppIcc95,bTseIsit97,bViswanathTit02}.
In a single-input single-output (\SISO) system, this scheme is known to be optimal in the sense of maximizing the sum rate \cite{bKnoppIcc95}.
Meanwhile, user unfairness can result from the asymmetric user fading statistics wherein a channel resource is likely to be dominated by strong users \cite{bViswanathTit02}.
To provide user fairness in addition to achieving multiuser diversity, fair schedulers employing a proportional fair or one-round-robin schemes are used \cite{bJorswieckEurasip09}.
The main idea of such fair schedulers is to schedule users on their own maximum/optimum channel \cite{bViswanathTit02,bJorswieckEurasip09}.

Frequency selectivity of a fading channel is usually due to resolvable multipaths in a channel which controls the degree of channel fluctuation in the frequency domain and provides frequency diversity benefits \cite{bLeeWiley93}.
While frequency selectivity complicates channel estimation, this form of diversity can be exploited by employing advanced techniques at a receiver such as maximal ratio combining (\MRC) or minimum mean squared estimation (\MMSE) \cite{bJakesTvt71,bMaTwc07}.
It improves the bit error rate (\BER) in single carrier systems \cite{bMaTwc07} and increases outage capacity in multicarrier systems such as orthogonal frequency division multiplexing (\OFDM) \cite{bAssaliniTvt09}.

In particular, for an \OFDM{} system operating in a channel with limited fluctuations, cyclic delay diversity (\CDD) was proposed to increase frequency selectivity and achieve the better \BER{} or outage performance \cite{bDammannGcom01,bBauchVtcf04,bAssaliniTvt09}.
This is an extension of conventional delay diversity in \cite{bSeshadriVtc93} to \OFDM{} systems.
Cyclic delay provides a mechanism to increase frequency selectivity by increasing the effective number of paths in the resulting channel.
Based on results in \cite{bDsYooTwc05NMSV} where it is shown that more frequency selective channels result in the lower \BER{}, it is advantageous to have larger cyclic delays in a channel when channel estimation is ideal \cite{bDammannGcom01}.
In \cite{bAssaliniTvt09}, the outage performance with respect to frequency selectivity was investigated showing that larger selectivity, as measured by the root mean square (\RMS) delay spread, leads to the better outage performance.
In \cite{bDsYooTwc05NMSV,bDsYooTwc05DC}, a new measure of frequency selectivity was proposed, \ie{} the inverse of the sum correlation of frequency components of a channel.
They showed that the measure correlates with \BER{} performance in a channel better than the conventional measure, the \RMS{} delay spread.

In \cite{bZhouTsp07,bChenTcom06}, the relation between multiuser diversity and spatial diversity using multiple antennas is explored in the flat fading channel context.
However, multiuser diversity has not been well studied with respect to the multipath channel, \ie{} frequency selectivity.
In \cite{bFlorenVtcs05}, the interaction between multiuser diversity and multipath diversity was studied when the scheduling unit block is the whole frequency band and when the maximum signal to noise ratio (\SNR) user scheduling is employed.
It was shown that the flat fading channel is the best in view of \SNR{}-based selection of the users.
However, if we consider a sub-block of the whole frequency band as a scheduling unit, as is the general scheme in orthogonal frequency division multiple access (\OFDMA) systems, and consider fair scheduling as well, we show that the flat fading channel is not the best because the lack of diversity between blocks is likely to decrease the sum rate.
Alternately, too large frequency selectivity is likely to decrease  the block average throughput, which also leads to a decrease of the sum rate indicating that there is an optimal interplay between multiuser diversity and frequency diversity.

In this paper, to understand the interplay between frequency selectivity and multiuser diversity, we investigate the effect of frequency selectivity on an \OFDMA{} multiuser system, where proportional fair scheduling is employed for user fairness.
We assume that the scheduling unit is a block of contiguous subcarriers.
As a measure of system performance, we choose the maximum of the block average throughput, and we show that this measure is a function of both intra-block and inter-block subcarrier correlation.
We develop approximate expressions to the maximum of the block average throughput of an arbitrary user, and use them to show that there exists an optimal frequency selectivity profile which maximizes multiuser diversity. Utilizing the insights from this study, we then show how \CDD{} techniques can be used to effectively control frequency selectivity.
We propose two techniques to optimally add frequency selectivity, \ie{} determine per-user optimal cyclic delay for \CDD{}, in a limited fluctuating channel.
We show that our techniques achieve the large gain compared to the standard \SISO{} technique and that the throughput is very close to the optimal sum rate possible with \CDD{}.

In summary, the paper has two main contributions.
First, we provide an analytical relationship between multiuser diversity and  frequency selectivity, and characterize optimal frequency selectivity.
Second, we develop two \CDD-based techniques to optimally control frequency selectivity in a given channel to maximize system throughput.

This paper is organized as follows.
In \Section{sSystemModel4FreqDiv}, we describe the channel and system model.
In \Section{sOptFreqSel}, we study the nature of the optimal frequency selectivity structure for maximizing the maximum of the block average throughput of an arbitrary user.
In \Section{sAddFreqSel}, we develop two \CDD-based techniques to control frequency selectivity of the channel by determining the proper value for the cyclic delay based on a power delay profile (\PDP{}) and an \RMS{} delay spread, respectively.
In \Section{sNumericalResults4FreqDiv}, we provide numerical results to support the theory developed.
They confirm the interplay between frequency selectivity and system throughput, and desirable frequency selectivity for maximizing throughput.
They also document the effectiveness of our \CDD-based techniques to add frequency selectivity.
We conclude in \Section{sConclusion4FreqDiv}.

\section{System Model}
\label{sSystemModel4FreqDiv}
\begin{figure}
       \centering
           \includegraphics[width=8.0cm]{\DetCdFigDir/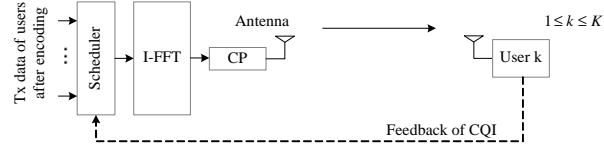}
           \caption{System block diagram of a multiuser \OFDMA{} system.}
           \label{fSysBldForFreqDiv}
   \end{figure}

We consider a single-input single-output (\SISO) complex Gaussian broadcast channel with one base station and $K$ users as shown in \Fig{fSysBldForFreqDiv}.
An \OFDMA{} system is assumed where $\Nsc$ and $T$ denote the length (in samples) and the time interval respectively of the \FFT{} (Fast Fourier Transform) used in the \OFDM{} system.
$\Nsc$ also equals the total number of subcarriers.
A frequency selective channel is assumed and the discrete time channel is given by
   \begin{equation}
       \label{eChannelForFreqDiv}
               h(t)= \sum_{m=1}^{L} \alpha_m h_m \delta \left( t- \tfrac{(m-1) T}{\Nsc} \right),
   \end{equation}
where $L$ is the number of paths, $\alpha_m$ is the average gain of path-$m$ (\ie{} $\sum_{m=1}^{L} \alpha_m^2 = 1$), and $h_m$ is the fading coefficient of path-$m$, which is modeled as $\CN (0,$$1)$, \iid{} in $m$.\footnote{$\CN (\mu,$$\sigma^2)$ denotes a circularly symmetric complex Gaussian distribution with mean $\mu$ and variance $\sigma^2$.}
The frequency response at subcarrier-$n$ is given by
   \begin{equation}
       \label{eHn}
       H_n = \sum_{m=1}^{L} \alpha_m h_m e^{-j \tfrac{2 \pi (m-1) n}{\Nsc}}, \quad 1 \leq n \leq \Nsc
   \end{equation}
Then, the received signal at subcarrier-$n$ satisfies the equation $Y_n = H_n X_n + W_n$, where $X_n$ is the transmitted symbol and $W_n$ is additive white Gaussian noise (\AWGN) with $\CN (0,\sigma_w^2)$.
The received \SNR{} on subcarrier-$n$, denoted by $\gamma_n$, is given by $\gamma_n={P |H_n|^2}/{\sigma_w^2}$, where $\Ex[|X_n|^2]=P$.
Based on the assumptions on $h_m$, the $H_n$'s are jointly Gaussian with the marginal density of $H_n$ being $\CN(0,1).$
The \SNR{} $\gamma_n$ follows a Gamma distribution $\Gcal(1,\frac{\sigma_w^2}{P})$.\footnote{$\Gcal(\alpha,\beta)$ denotes a Gamma distribution with \PDF{} \cite{bGarciaAw94}, $f_{\gamma_n}(\gamma) = \frac{\beta^\alpha}{\Gamma(\alpha)} \gamma^{\alpha-1} e^{-\beta \gamma}$, where $\Gamma(\alpha) = \int_0^\infty t^{\alpha-1} e^{-t} dt$.}

\begin{figure}
       \centering
           \includegraphics[width=5.0cm]{\DetCdFigDir/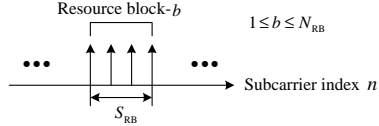}
           \caption{Contiguous subcarrier grouping.}
           \label{fRbStructure}
   \end{figure}

In a multiuser system, the throughput is larger when the resource allocation is flexible and has high granularity, \eg{} assignment at the individual subcarrier level.
However, the complexity and feedback overhead can be prohibitive calling for simpler approaches.
In our work, the overall $\Nsc$ subcarriers are grouped into $\Nrb$ number of resource blocks ($\RB$), and each block contains contiguous $\Srb$ subcarriers as in \Fig{fRbStructure}, where $\Nsc$$=$$\Nrb$ $\times$ $\Srb$.
The assignment is done at the block level, \ie{} a resource block is assigned to a user.
The block size ($\Srb$) is assumed to be known and in practice can be determined at the medium access control (\MAC) layer taking into account the number of users.
A measure used for resource allocation is the block average throughput $C_b$, which for block-$b$ of a user is given by
   \begin{equation}
       \label{eCb}
           C_b \triangleq \frac{1}{\Srb} \sum_{{ n = 1 + (b-1)\Srb }}^{b \Srb} \log_2 (1+\gamma_n).
   \end{equation}
For scheduling purposes, it is assumed that each user feeds back the ordered best-$\NFb$ block average throughput values ($C_b$) together with the block indices to the transmitter.
The feedback is assumed to be error-free and with no-delay.

\subsection{Proportional fair scheduling}
\label{sPropFairScheduling}
To prevent a user with a good channel from being allocated a disproportionate number of resource blocks, the transmitter schedules users employing a proportional fair scheme based on the feedback information provided by them \cite{bViswanathTit02}.
Since we have $\Nrb$ blocks and a grouping scheme is used, there are $\Nrb$ steps in the assignment of blocks to users at each time $t$.
In this approach, user $k_\ell^*$ is scheduled to a block $b_\ell^*$ at $\ell$-th assignment in time $t$ as follows $(1 \leq \ell \leq \Nrb )$:
    \begin{equation}
       \label{ePf}
           (k_\ell^*,b_\ell^*) = \arg \max_{k \in \textrm{\{all users\}}} \; \max_{b \in \textrm{\{remaining blocks}\}} \RpfmKb \big(t+\tfrac{\ell}{\Nrb} \big),
    \end{equation}
where $\RpfmKb$ denotes the proportional fair metric in block-$b$ of user-$k$, and is given by $C_{k,b} / \RavgK$, where $C_{k,b}$ denotes the block average throughput of user-$k$ as per \eqref{eCb} and $\RavgK $ denotes the average throughput of user-$k$.
Once a user is scheduled in $\ell$-th assignment, the average throughputs for all users are updated in the following manner.
\begin{equation}
   \label{ePfm}
           \RavgK\big(t+\tfrac{\ell}{\Nrb} \big)= \begin{cases}
               \left(1 - \frac{1}{t_c} \right) \RavgK \big(t+\tfrac{\ell-1}{\Nrb} \big) + \frac{1}{t_c} \; C_{k_\ell^*,b_\ell^*}, \hspace{0.1cm} k=k_\ell^* \\
               \left(1 - \frac{1}{t_c} \right) \RavgK \big(t+\tfrac{\ell-1}{\Nrb} \big), \hspace{1.4cm} k \neq k_\ell^*
           \end{cases}.
\end{equation}
Here $t_c$ is the length of scheduling window \cite{bViswanathTit02}.
The sum rate of the system is given by
\begin{equation}
   \label{eRsumForFreqDiv}
           \hspace{-3.0cm} \Rsum = \frac{1}{\Nrb} \sum_{\ell=1}^{\Nrb} \Ex [ C_{k_\ell^*, b_\ell^*} ].
\end{equation}

This sort of proportional fair scheduling is highly likely to schedule users to their own maximum block with the largest block average throughput across the entire frequency band for the selected user \cite{bViswanathTit02}.
This situation becomes more lively when the number of users increases as well as when the number of feedback is one (\ie{} $\NFb=1$).
{\it This leads us to assume that the sum rate gain (multiuser diversity) is directly related to maximizing the maximum of the block average throughput of an arbitrary user in the entire band (\ie{} maximizing $\max_b C_b$).} We now focus on how frequency selectivity affects the maximum of the block average throughput in \OFDMA{} multiuser scheduling systems.

\section{Optimal frequency selectivity that maximizes the maximum of the block average throughput ($\ExMaxCb$)}
\label{sOptFreqSel}
When we consider a system without feedback, frequency selectivity improves the bit error rate \cite{bDsYooTwc05NMSV} or outage performance \cite{bAssaliniTvt09}.
However, when we consider a scheduling system with feedback based on a block of subcarriers, frequency selectivity will not always improve the sum rate.
To study this more analytically, we first examine a measure for frequency selectivity.
We then provide an approach to investigate the relation between the maximum of the block average throughput $\Ex[\max_b C_b]$ and frequency selectivity of a channel.
Finally, we show that there exists optimal frequency selectivity that maximizes the maximum of the block average throughput.
For this purpose we define useful functions below, which are also shown in \Table{tNoteSumForCC} for easy reference.

\begin{table}[h]
    \centering
    \caption{Notation summary of useful functions.}
    \label{tNoteSumForCC}
    \begin{tabular}[c]{|c|p{\TableColumnSizeOne}|}
        \hline
            Notation
            &
            Definition
            \\
            \hline

            \raisebox{-0.1ex}[0pt]{$\RhoSc(|\Delta_n|)$}
            &
            \CC-\SC: Correlation coefficient of the \SNR{} between two subcarriers apart by $\Delta n$.
            \\
            \hline

            \raisebox{-1.1ex}[0pt]{$\SumSc(r,|\Delta_b|,\Srb)$}
            &
            \begin{tabular}[c]{@{} p{\TableColumnSizeOne}}
                Sum of $\RhoSc(|\Delta_n|)$ between every possible two subcarriers each in two blocks apart by $\Delta_b$.\\
                \hline
                $\SumSc(1,0,\Srb)$: {\it Intra-block sum correlation}.\\
                $\SumScInv$: {\it Frequency selectivity measure}.
            \end{tabular}
            \\
            \hline

            \raisebox{-0.0ex}[0pt]{$\RhoRb(|\Delta_b|)$}
            &
            \CC-\RB: Correlation coefficient of the block average throughput between two blocks apart by $\Delta_b$.
            \\
            \hline

            \raisebox{-1.0ex}[0pt]{$\SumRb(\Srb)$}
            &
            \begin{tabular}[c]{@{} p{\TableColumnSizeOne}}
                {\it Inter-block sum correlation}: Sum of $\RhoRb(|\Delta_b|)$ between every possible two blocks in the whole band.\\
                \hline
                $\SumRbInv$: Effective number of blocks.
            \end{tabular}
            \\
            \hline

    \end{tabular}
\end{table}

\subsection{Characterization of frequency selectivity of a channel}
\label{sCharFreqSelOfCh}
\subsubsection{Some of useful functions}
\label{sDefineUsefulFncForFreqSel}
Since frequency selectivity of a channel indicates similarity or difference between subcarriers, it can be described by the statistical correlation property between subcarriers.
As a basic measure characterizing frequency selectivity, we first define the correlation coefficient of the \SNR{} between two subcarriers indexed by $\NOne$ and $\NTwo$ (\CC-\SC) as \cite{bGarciaAw94}
    \begin{equation}
        \label{eRhoSnrSc}
            \RhoSc(|\Delta_n|) \triangleq \frac{\Cov(\gamma_\NOne,\gamma_\NTwo)}{\sqrt{\Var[\gamma_\NOne]} \sqrt{\Var[\gamma_\NTwo]}}
    \end{equation}
where $\Delta_n = \NTwo - \NOne$ and `SC' stands for the `subcarrier'.
`$\Cov$' and `$\Var$' denote covariance and variance respectively.
It is shown in \Append{sDeriveRhoSc} that for the channel in \eqref{eChannelForFreqDiv}, we have
    \begin{equation}
        \label{eRhoSnrScDerive}
            \RhoSc(|\Delta_n|) = |\Cov(H_\NOne,H_\NTwo)|^{2},
    \end{equation}
where it can be shown from \eqref{eHn} that
    \begin{equation}
        \label{eCovH}
             \Cov(H_\NOne, H_\NTwo) = \sum_{m=1}^{L} \alpha_m^2 e^{ -j \tfrac{2\pi}{\Nsc} (m -1) (\NTwo - \NOne) }.
    \end{equation}
We note from \eqref{eRhoSnrScDerive} and \eqref{eCovH} that $\RhoSc$ is a function of $|\Delta_n|$, and that $\RhoSc$ is periodic with a period $\Nsc$, \ie{} $\RhoSc(|\Delta_n|) = \RhoSc(|\Delta_n - \Nsc|)$.
By the nonnegativity of $\RhoSc(|\Delta_n|)$ in \eqref{eRhoSnrScDerive} and the magnitude property of the correlation coefficient \cite{bGarciaAw94} (\ie{} $-1 \leq \RhoSc(|\Delta_n|) \leq 1$), we find that $0 \leq \RhoSc(|\Delta_n|) \leq 1$.

Since the scheduling unit is a subcarrier block in \OFDMA{} systems, we need to know frequency selectivity between blocks.
To state the correlation between blocks, we define the sum of correlation coefficients of the \SNR{} between subcarriers in each of the two blocks indexed by $\BOne$ and $\BTwo$ as
    \begin{equation}
        \label{eSumSc}
                {{\SumSc(r,|\Delta_b|,\Srb) \triangleq \frac{1}{\SrbSq}}}
                \SumScInTwoRb
                [{{\RhoSc(|\Delta_n|)}}]^r.
    \end{equation}
where $\Delta_b = \BTwo - \BOne$ and $r$ is a free parameter related to the order of expansion of $\log_2(1+\gamma_n)$ in \eqref{eCb}.
In our analysis, $r=1$ for the measure of frequency selectivity in \eqref{eSumScOneZero}.
The case that $r=2$ is shown in \eqref{eVarCbSoe} of \Append{sDeriveStatCb} for the second order approximation of the variance of the block average throughput.
We note in \eqref{eSumSc} that sum is over every possible combination of subcarriers from blocks $\BOne$ and $\BTwo$ respectively.
By replacing the summation index, we can rewrite \eqref{eSumSc} as
    \begin{equation}
        \label{eSumSc2}
                {{\SumSc(r,|\Delta_b|,\Srb) = \frac{1}{\SrbSq}}}
                \sum_{\NOne'=1}^{\Srb} \sum_{\NTwo'=1}^{\Srb}
                [{{\RhoSc(|\Delta_b \Srb + \NTwo' - \NOne'|)}}]^r,
    \end{equation}
where we can verify that $\SumSc$ depends on $|\Delta_b|$ utilizing \eqref{eRhoSnrScDerive} and \eqref{eCovH}.
For $\SumSc(r,|\Delta_b - \Nrb|,\Srb)$, we note in the argument of $\RhoSc$ in \eqref{eSumSc2} that $|(\Delta_b - \Nrb)\Srb + \NTwo' - \NOne'| = |(\Delta_b )\Srb + \NTwo' - \NOne' - \Nsc| \stackrel{(a)}{\equiv} |(\Delta_b )\Srb + \NTwo' - \NOne'|$, where the last equivalence $(a)$ follows from the periodicity of $\RhoSc$.
Thus, we can find that $\SumSc$ is also periodic with a period of $\Nrb$, \ie{} $\SumSc(r,|\Delta_b|,\Srb)$ = $\SumSc(r,|\Delta_b - \Nrb|,\Srb)$.

As a special case, for the same block ($\Delta_b=0$) and for the first order ($r=1$), we have
    \begin{equation}
        \label{eSumScOneZeroOri}
                {{\SumSc(1,0,\Srb) = \frac{1}{\SrbSq}}}
                \sum_{\NOne=1}^{\Srb} \sum_{\NTwo=1}^{\Srb}
                {{\RhoSc(|\Delta_n|)}}.
    \end{equation}
Since this sum is for subcarriers within an identical block, it is referred to as {\it intra-block sum correlation}.
Since $0 \leq \RhoSc(|\Delta_n|) \leq 1$ and $\RhoSc(0)=1$, we find from \eqref{eSumScOneZeroOri} that $ \frac{1}{\Srb} \leq \SumSc(1,0,\Srb) \leq 1$ where the minimum is for a channel with independent subcarriers, and the maximum is for a flat channel.

\subsubsection{Measure of frequency selectivity of a channel}
\label{sSumScInv}
As one measure to characterize frequency selectivity, the inverse of the intra-block sum correlation in \eqref{eSumScOneZeroOri} for the whole band (\ie{} $\Srb = \Nsc$) is used in \cite{bDsYooTwc05NMSV,bDsYooTwc05DC}.
Considering $\RhoSc(|\Delta_n|) = \RhoSc(|\Delta_n - \Nsc|)$, we have from \eqref{eSumScOneZeroOri} that $\SumSc(1,0,\Nsc)=\frac{1}{\Nsc} \sum_{n=0}^{\Nsc-1} \RhoSc(n)$.
Thus, its inverse is given by
    \begin{equation}
        \label{eSumScOneZero}
            \SumScInv= \frac{1}{\frac{1}{\Nsc} \sum_{n=0}^{\Nsc-1} \RhoSc(n)}.
    \end{equation}
We note in \eqref{eSumScOneZero} that the frequency selectivity is inversely proportional to the average correlation coefficient in the whole band.
This agrees with the intuition that an increase of frequency selectivity makes a channel more fluctuating, which leads to a decrease of the correlation coefficient of the \SNR{} between subcarriers \cite{bDsYooTwc05NMSV} and the sum correlation in \eqref{eSumScOneZeroOri}, and an increase of its inverse $\SumScInv$ in \eqref{eSumScOneZero}.
Thus, we regard large frequency selectivity (\ie{} $\SumScInv$) as the small intra-block sum correlation and vice versa throughout the paper.

In addition to being used as a measure for frequency selectivity, $\SumScInv$ is used as the effective number of paths in a channel \cite{bDsYooTwc05NMSV,bDsYooTwc05DC,bBarriacTcom04,bKimTbc09}.
Providing some intuition about this relationship, we first check the following equation from \cite[(11)]{bAssaliniTvt09} and \cite[(9)]{bKimTbc09}.
    \begin{equation}
        \label{eSumScInv}
                \SumScInv = \frac{\Var[\gamma_1]}{\Var \left[ \frac{1}{\Nsc} \sum_{n=1}^\Nsc \gamma_n \right]} = \frac{1}{\sum_{m=1}^{L} \alpha_m^4}.
    \end{equation}
This indicates the effective number of paths in a channel when the gains of the paths are made equal (\ie{} $\SumScInv=L \textrm{ when } \alpha_m= \sqrt{1/L}$).
For example, for two equal paths ($\alpha_1 = \alpha_2 = {\sqrt{{1}/{2}}}$), $\SumScInv$ is exactly 2.
However, when $\alpha_1 = \sqrt{{2}/{3}}$ and $\alpha_2 = \sqrt{{1}/{3}}$, $\SumScInv$ is $\frac{9}{5}$.
The conventional diversity order for these two cases is the same value of 2 since diversity is a high \SNR{} measure \cite[(9.3)]{bTseCup05}.
However, $\SumScInv$ can differentiate.

When frequency diversity provided by the multipaths is exploited in the frequency domain of \OFDM{} systems, the order of frequency diversity, \ie{} the effective number of independent subcarriers, is the same as the effective number of paths.
Thus, the effective number of independent subcarriers is the same as $\SumScInv$.
For example, suppose that all the subcarriers are completely correlated.
Then, $\RhoSc(|\Delta_n|)=1$ for any $\Delta_n=\NTwo-\NOne$.
Thus, $\SumScInv= 1$ in \eqref{eSumScOneZero}.
Since all subcarriers have the same value in each channel realization, the frequency diversity order is one and the effective number of independent subcarriers is one.
Thus, the effective number of independent subcarriers matches with $\SumScInv$.
For another example of all independent subcarriers, we have $\RhoSc(|\Delta_n|)=0$ for different subcarriers ($\Delta_n \neq 0$) and $\RhoSc(|\Delta_n|)=1$ for itself ($\Delta_n=0$).
Thus, $\SumScInv = \Nsc$, which is the same as the effective number of independent subcarriers.

\subsection{Development of the relation between $\ExMaxCb$ and frequency selectivity}
\label{sApprMaxCb}
\subsubsection{Function definitions for inter-block frequency selectivity and effective number of resource blocks}
\label{sRhoRb}
As we briefly mentioned in \Section{sDefineUsefulFncForFreqSel}, we need to characterize inter-block frequency selectivity since we consider a block-based \OFDMA{} system.
As a basic measure for this purpose, we define the correlation coefficient of the block average throughput between two blocks indexed by $\BOne$ and $\BTwo$ (\CC-\RB) as \cite{bGarciaAw94}
    \begin{equation}
        \label{eRhoCapRb}
            \RhoRb(|\Delta_b|) \triangleq \frac{\Cov(C_\BOne,C_\BTwo)}{\sqrt{\Var[C_\BOne]} \sqrt{\Var[C_\BTwo]}}
    \end{equation}
where `RB' stands for the `resource block' and we follow the same notations in \Section{sDefineUsefulFncForFreqSel}.
For the first order approximation of $C_b$, it is shown in \Append{sDeriveStatCb} that we have
    \begin{equation}
        \label{eRhoCapRbFoa}
                \RhoRb(|\Delta_b|) = \frac{\SumSc(1,|\Delta_b|,\Srb)}{\SumSc(1,0,\Srb)}
    \end{equation}
where we verify that this is a function of $|\Delta_b|$.
We can easily verify that $0 \leq \RhoRb(|\Delta_b|) \leq 1$ from the nonnegativity of $\SumSc$ in \eqref{eSumSc} and the magnitude property of the correlation coefficient and that $\RhoRb(|\Delta_b|) = \RhoRb(|\Delta_b - \Nrb|)$ from the periodicity of $\SumSc(1,|\Delta_b|,\Srb)$ in \eqref{eSumSc}.

\label{sEffRbNum}
In the same line of context for \eqref{eSumScOneZeroOri}, we define the sum of correlation coefficients of the block average throughput between every possible two blocks in the whole band as
    \begin{equation}
        \label{eSumRb}
                \SumRb(\Srb) \triangleq \frac{1}{\NrbSq} \SumRbInAll \RhoRb(|\Delta_b|).
    \end{equation}
Since this sum is for all the blocks in the whole band, it is referred to as {\it inter-block sum correlation}.
From the periodicity of $\RhoRb(|\Delta_b|)$, \eqref{eSumRb} is reduced to
    \begin{equation}
        \label{eSumRbRed}
            \SumRb(\Srb)= \frac{1}{\Nrb} \sum_{b=0}^{\Nrb-1} \RhoRb(b).
    \end{equation}
We note that the inter-block sum correlation is the average correlation among blocks in the whole band.

The discussion about effective number of subcarriers (\ie{} $\SumScInv$) in \Section{sSumScInv} motivates defining the effective number of independent blocks as $\SumRbInv$, which is the inverse of the inter-block sum correlation in \eqref{eSumRbRed}.
We can verify from \eqref{eSumRbRed} that $1 \leq \frac{1}{\SumRb(\Srb)} \leq \Nrb$ where the minimum is for a flat channel (\ie{} $\RhoRb(|\Delta_b|)=1$ for all $\Delta_b$), and the maximum is for a channel with independent blocks (\ie{} $\RhoRb(|\Delta_b|)=0$ for $\Delta_b\neq0$ and $\RhoRb(|\Delta_b|)=1$ for $\Delta_b=0$).
In these both extreme cases of frequency selectivity of a channel, we can easily verify that the effective number $\SumRbInv$ is the same as the number of independent blocks.
Noting \eqref{eSumScInv} and the analogy between $\SumScInv$ and $\SumRbInv$, we can verify for the first order approximation of $C_b$ that
    \begin{equation}
        \label{eSumRbInv}
                \SumRbInv = \frac{\Var[\sum_{n=1}^\Srb \frac{\gamma_n}{\Srb} ]}{\Var \left[ \frac{1}{\Nrb} \sum_{b=1}^\Nrb \left( \sum_{n=1 + (b-1)\Srb}^{b \Srb} \frac{ \gamma_n}{\Srb}  \right) \right]}.
    \end{equation}
Considering from \eqref{eSumScInv} that $\frac{1}{\SumSc(1,0,\Srb)} = \frac{\Var[\gamma_1]}{\Var \left[ \frac{1}{\Srb} \sum_{n=1}^\Srb \gamma_n \right]}$, we have from \eqref{eSumScInv} and \eqref{eSumRbInv} as
    \begin{equation}
        \label{eSumScSumRb}
                \SumScInv = \frac{1}{\SumSc(1,0,\Srb)} \times \SumRbInv.
    \end{equation}
This gives the idea that the effective number of subcarriers in the whole band at the left-hand side is the same as the product of the effective number of blocks in the whole band and the effective number of subcarriers in each effective block at the right-hand side.

\subsubsection{Approximations of $\ExMaxCb$ and optimality in frequency selectivity that maximizes $\ExMaxCb$}
Suppose that we have $N$ \iid{} random variables of $X_i$ ($1 \leq i \leq N$) and that $Y$ is the maximum of $X_i$'s.
That is, $Y = \max_{1 \leq i \leq N} X_i$.
When a probability density function (\PDF{}) of $X_i$ is not given in a closed-form,\footnote{For example, suppose that $X_i$ is the sum of dependent random variables, say $Z_j$ for $1 \leq j \leq m$ and that we know only \PDF{} of $Z_j$ and their correlation.
It is usually hard or intractable to obtain the \PDF{} of $X_i$.
However, we can compute $\Ex[X_i]$ and $\Var[X_i]$.} it is usually not tractable to compute $\Ex[Y]$.
However, we can obtain some insight about the relation between $\Ex[Y]$ and $\{ \Ex[X_i],\Var[X_i],N \}$ from a simple upper bound of the order statistics \cite{bDavidJwas04}
    \begin{equation}
        \label{eOsUb}
                \Ex[Y] \leq \Ex[X_i] + \frac{N-1}{\sqrt{2N-1}} \sqrt{\Var[X_i]}.
    \end{equation}
While this bound is good for the small $N$, it becomes loose when $N$ becomes larger.
In a special case that $X_i$ is Gaussian random variable, the weak law of large number gives an approximation of $\Ex[Y]$ as \cite{bHochwaldTit04}
    \begin{equation}
        \label{eGaUb}
                \Ex \left[ Y \right] \simeq \Ex[X_i] + \sqrt{2 \Var[X_i] \ln N}.
    \end{equation}
This approximation is better for large $N$.
We note in \eqref{eOsUb} and \eqref{eGaUb} that the expectation of the maximum of $X_i$ (\ie{} $\Ex[Y]$) increases with two moments of $X_i$ (\ie{} $\Ex[X_i]$ and $\Var[X_i]$) and the number of $X_i$ (\ie{} $N$).

From the assumption in \Section{sPropFairScheduling} that the sum rate gain (multiuser diversity) is directly related to the maximum of the block average throughput by the proportional fair scheduling, we focus on approximating $\ExMaxCb$.
Using $\Ex[C_b]$ in \eqref{eExpCb} and $\Var[C_b]$ in \eqref{eVarCbFoe} in \Append{sDeriveStatCb}, we can approximate $\ExMaxCb$ by replacing $N$ in \eqref{eOsUb} with the effective number of blocks $\SumRbInv$ in \eqref{eSumRbRed} as
    \begin{equation}
        \label{eOsApprExMaxCb}
                \ExMaxCb \leq E_1 + \tfrac{\left(\SumRbInv - 1 \right)\sqrt{ \SumSc(1,0,\Srb)}}{\sqrt{ \frac{2}{\SumRb(\Srb)} - 1}} \sqrt{V_1}.
    \end{equation}
where $E_1 = \Ex[\log_2(1+\gamma_1)]$ and $V_1 = \frac{\Var[\gamma_1]}{\{ (1 + \Ex[\gamma_1]) \ln 2 \}^2}$ for notational simplicity.

In \cite{bBarriacTcom04,bAssaliniTvt09}, Gaussian approximation of $C_b$ in \eqref{eCb} is suitable for identically distributed $\gamma_n$ when the system bandwidth is large.
Since we consider a block of wideband systems, we can apply this theorem for the reasonable block size.
We will show the justification of this assumption in the numerical results.
Thus, we can assume that $C_b$ follows $\Ncal(\Ex[C_b],\Var[C_b])$.\footnote{$\Ncal (\mu,$$\sigma^2)$ denotes a Gaussian distribution with mean $\mu$ and variance $\sigma^2$.}
Using \eqref{eExpCb}, \eqref{eVarCbFoe}, and the effective number of blocks $\SumRbInv$ in \eqref{eSumRbRed}, we can approximate $\ExMaxCb$ using the relation in \eqref{eGaUb} as
    \begin{equation}
        \label{eGaApprExMaxCb}
                \ExMaxCb \simeq E_1 + \sqrt{\SumSc(1,0,\Srb) \ln \tfrac{1}{\SumRb(\Srb)}} \sqrt{ 2 V_1}.
    \end{equation}
We note that the second order expansion of $\Var[C_b]$ in \eqref{eVarCbSoe} in \Append{sDeriveStatCb} can be used in \eqref{eOsUb} and \eqref{eGaUb} to obtain more accurate approximations.

From \eqref{eOsApprExMaxCb} and \eqref{eGaApprExMaxCb}, we can note two important facts when a marginal distribution of the \SNR{} ($\gamma_n$) is fixed.
First, the maximum $C_b$ of a user increases with $\SumSc(1,0,\Srb)$, {\it intra-block sum correlation}.
This means that subcarriers within a block should be highly correlated to increase the maximum of $C_b$.
Thus, the flat fading is the best case in this view.
On the other hand, the maximum $C_b$ of a user increases with $\SumRbInv$, the inverse of {\it inter-block sum correlation}.
This means that blocks should be lowly correlated to increase the maximum of $C_b$.
Thus, frequency selective fading with larger $\SumRbInv$ is preferred in this view.
Thus, for larger $\ExMaxCb$, we need the large intra-block sum correlation and the small inter-block sum correlation.
As the number of paths in a channel increases, $\SumSc(1,0,\Srb)$ decreases but $\SumRbInv$ increases.
Thus, we note that there exists a trade-off between these two factors, \ie{} intra-block sum correlation and inter-block sum correlation.

To find an optimality of frequency selectivity for $\ExMaxCb$, let us look at \\ $\Ecal \triangleq \sqrt{\SumSc(1,0,\Srb) \ln \tfrac{1}{\SumRb(\Srb)}}$ in \eqref{eGaApprExMaxCb}.
We note that $\Ecal$ indicates the additional gain of expectation by the maximum selection compared to the individual one (\ie{} $E_1$ in \eqref{eGaApprExMaxCb}).
We consider $\Ecal$ for three types of channels.
One is a flat channel (CH\_A), other is a channel with independent subcarriers (CH\_B) and another is an ideal channel which is flat within a block and mutually independent between blocks (CH\_C).
Following the discussion in \Section{sSumScInv} and \Section{sEffRbNum}, we have $\SumSc(1,0,\Srb)$ and $\SumRbInv$ in \Table{tExMaxCb} for each channel.
We note that $\SumScInv$ can be computed from \eqref{eSumScSumRb}.
From the table, we can find that CH\_C has the largest $\Ecal$, which leads to the largest $\ExMaxCb$ in \eqref{eGaApprExMaxCb}.
However, frequency selectivity of CH\_C is less than CH\_B (a channel with independent subcarriers).
We note that both extreme cases of a channel, \ie{} flat or fully independent, are not good for maximizing $\ExMaxCb$.
This tells us that there may exist optimal frequency selectivity between a flat channel and an independent channel.
Further, a channel with optimal selectivity should be like CH\_C, \ie{} as flat as possible inside a block and as independent as possible among blocks, which complies with the observation in \cite{bKhanVtcf06,bHimayatGcom07}.
    \begin{table}[h]
        \centering
        \begin{threeparttable}
            \centering
            \caption{Comparison of $\ExMaxCb$ for three types of channels.}
            \label{tExMaxCb}
            \begin{tabular}[c]{|p{6.0cm}|p{2.0cm}|p{2.0cm}|p{2.0cm}|}
                \hline
                    Channel type    &   CH\_A   &   CH\_B   &   CH\_C   \\
                \hline
                \hline
                    $\SumSc(1,0,\Srb)$  &   1   &   $\tfrac{1}{\Srb}$    &   1   \\
                \hline
                    $\SumRbInv$  &   1   &   $\Nrb$    &   $\Nrb$   \\
                \hline
                    Frequency selectivity=$\tfrac{1}{\SumSc(1,0,\Nsc)}$   &   1   &   $\Nsc$  &   $\Nrb$  \\
                \hline
                    $\Ecal=\sqrt{\SumSc(1,0,\Srb) \ln \tfrac{1}{\SumRb(\Srb)}}$ &   $0$ &   $\sqrt{\tfrac{\ln \Nrb}{\Srb}}$   &   $\sqrt{\ln \Nrb}$   \\
                \hline
            \end{tabular}
            \begin{tablenotes}
                \item[1] CH\_A denotes a flat channel.
                    CH\_B denotes a channel with independent subcarriers.
                    CH\_C denotes a channel which is flat within a block and mutually independent between blocks.

                \item[2] $\SumSc(1,0,\Srb)$ determines $\Var[C_b]$ in \eqref{eVarCbFoe} and $\SumRbInv$ represents the effective number of blocks.
                \item[3] $\tfrac{1}{\SumSc(1,0,\Nsc)}$ can be computed from \eqref{eSumScSumRb} and represents frequency selectivity.
                \item[4] $\sqrt{\SumSc(1,0,\Srb) \ln \tfrac{1}{\SumRb(\Srb)}}$ is from \eqref{eGaApprExMaxCb} and related to $\ExMaxCb$.
            \end{tablenotes}
        \end{threeparttable}
    \end{table}

In an open loop diversity system without feedback, the more frequency selective channel with low correlation between subcarriers is preferred to improve outage property \cite{bAssaliniTvt09} or the \BER{} \cite{bDsYooTwc05NMSV}.
However, we note from the above that there exists optimal frequency selectivity, \ie{} an optimal correlation in the frequency domain, that maximizes the maximum of $C_b$ for a scheduling system.

Although we cannot reduce frequency selectivity for a given channel, we can increase frequency selectivity using a cyclic delay diversity technique.
In \Section{sAddFreqSel}, we propose a technique regarding how much selectivity should be added to maximize $\ExMaxCb$ in a channel with low selectivity.

\section{Optimal addition of frequency selectivity using cyclic delay diversity}
\label{sAddFreqSel}
In the previous section, we noted in \eqref{eOsApprExMaxCb} and \eqref{eGaApprExMaxCb} that there exists optimal frequency selectivity in maximizing $\ExMaxCb$.
The question we consider in this section is how much more channel selectivity should be added to maximize $\ExMaxCb$ when we are given a limited fluctuating channel.
One method to increase the number of paths in a channel is to use multiple transmit antennas.
By sending the same signal in different antennas at the different time, we have an equivalent channel with more paths.
For example, suppose that we have two transmit antennas each with flat fading and equal power.
If we add a delay by one symbol time at the second transmit antenna, the equivalent channel at a receiver has two equal paths separated by one symbol time.
This sort of delay diversity was first proposed in the single carrier system \cite{bSeshadriVtc93} and later for \OFDM{} system in the name of cyclic delay diversity (\CDD) \cite{bDammannGcom01}.
Since cyclic delay in \CDD{} determines frequency selectivity of the equivalent channel, we focus on how large cyclic delay we need to choose to maximize $\ExMaxCb$.

\begin{figure}
        \centering
            \includegraphics[width=\FigSizeNew]{\DetCdFigDir/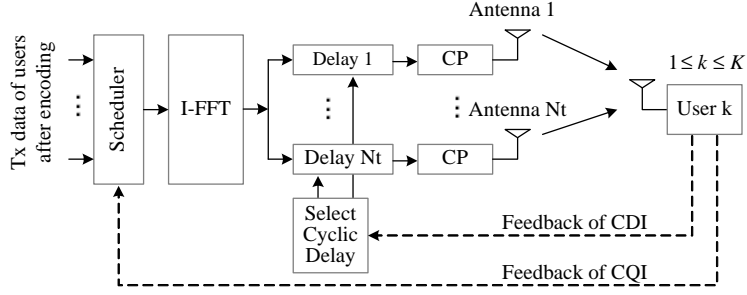}
            \caption{System block diagram of cyclic delay diversity (\CDD).}
            \label{fSysBldCdd}
    \end{figure}

Let $\NTx$ denote the number of transmit antennas in \Fig{fSysBldCdd}.
Let $D_i$ denote a cyclic delay in Tx antenna-$i$ and let $\Dvec=[D_1,\cdots,D_{\NTx}]^T$.
We note that $D_i$ has an integer value within $[0,\Nsc-1]$.
We follow the same notation in \eqref{eChannelForFreqDiv} except for adding an index $i$ to denote the transmit antenna.
Then, the discrete time channel equation is given by
    \begin{equation}
        \label{eCddChannel}
                \hCdd(t)= \sum_{i=1}^{\NTx} \sum_{m=1}^{L_i} \frac{\alphaim \him }{\sqrt{\NTx}} \; \delta \left( t- \tfrac{(m+D_i-1)_\Nsc T}{\Nsc} \right)
    \end{equation}
where $(\cdot)_\Nsc$ denotes modulo-$\Nsc$ operation.
Without loss of generality, we assume that $D_1$ is zero as in \cite{bBauchTwc06}.
We assume that $\him$ is \iid{} in $i$ and $m$.

Noting that $\Hin$ denotes a frequency response at subcarrier-$n$ in Tx antenna-$i$, we have the frequency response of \CDD{} at subcarrier-$n$ from \eqref{eCddChannel} as
    \begin{equation}
        \label{eCddFreqResp}
            \HnCdd = \sum_{i=1}^{\NTx} \frac{ \Hin }{\sqrt{\NTx}} \: e^{ -j \frac{2 \pi}{\Nsc} D_i n }.
    \end{equation}
Since $\HnCdd$ is linear combination of independent $\Hin$'s following $\CN(0,1)$ in \Section{sSystemModel4FreqDiv}, we can find that $\HnCdd$ follows $\CN(0,1)$ as well.

\subsection{Determination of cyclic delay from approximation of $\ExMaxCb$}
\label{sDetCdFromApprs}
Let $\RhoScCdd$ and $\RhoRbCdd$ denote the correlation coefficient for \CDD{} of the \SNR{} as in \eqref{eRhoSnrSc} and of $C_b$ as in \eqref{eRhoCapRb} respectively.
Let $\SumScCdd$ and $\SumRbCdd$ denote sum of $\RhoScCdd$ as in \eqref{eSumSc} and $\RhoRbCdd$ as in \eqref{eSumRb} respectively.
We can see that these values will be changed when we change cyclic delay because the channel delay profile (\PDP) is changed from \eqref{eChannelForFreqDiv} into \eqref{eCddChannel}.
For a given channel, these values will be a function of $\Dvec$, which will be shown later.
Let $\DPerUserVec$ denote optimal cyclic delay that maximizes $\ExMaxCb$ of an arbitrary user and $\DSumRateVec$ optimal cyclic delay that maximizes the sum rate.
Then problem we focus is to find $\DPerUserVec$ and to compare it to $\DSumRateVec$.
Further, we look at how much gain in the sum rate is achieved by this addition of frequency selectivity (\ie{} $\DPerUserVec$ or $\DSumRateVec$).

Using the approximations for $\ExMaxCb$ in \eqref{eOsApprExMaxCb} and \eqref{eGaApprExMaxCb}, we can find $\DPerUserVec$ in two ways as following.
    \begin{equation}
        \label{eDPerUserOsApprExMaxCb}
                \DPerUserVec = \arg \max_{\Dvec} \Big[ E_1 + \tfrac{\left(\frac{1}{\SumRbCdd(\Srb)} - 1 \right)\sqrt{ \SumScCdd(1,0,\Srb)}}{\sqrt{ \frac{2}{\SumRbCdd(\Srb)} - 1}} \sqrt{V_1} \Big].
    \end{equation}
    \begin{equation}
        \label{eDPerUserGaApprExMaxCb}
                \DPerUserVec = \arg \max_{\Dvec} \Big[ E_1 + \sqrt{\SumScCdd(1,0,\Srb) \ln \tfrac{1}{\SumRbCdd(\Srb)}} \sqrt{ 2 V_1} \Big] .
    \end{equation}
We note that we can omit $E_1$ and $V_1$ in both equations because the distribution of $\HnCdd$ is $\CN(0,1)$ and its statistics are not affected by $\Dvec$.

\subsubsection{Derivation of statistics of \CDD}
\label{sDeriveStatOfCdd}
Using the bilinear property of covariance \cite{bCovUrl}, we have from \eqref{eCddFreqResp} as
    \begin{equation}
        \label{eCovCddHOri}
             \Cov(\HnOneCdd, \HnTwoCdd) = \sum_{i=1}^{\NTx} \frac{\Cov(\HinOne,\HinTwo)}{\NTx} \; e^{ -j \tfrac{2\pi}{\Nsc} D_i (\NTwo - \NOne) }.
    \end{equation}
where $\Cov(\HinOne,\HinTwo)$ denotes covariance of \SISO{} channel at Tx antenna-$i$ in \eqref{eCovH}.
Note that covariance depends on $\Dvec$ as well as $\Delta_n$ (\ie{} $\NTwo-\NOne$).
Using this and following the same procedure in \Append{sDeriveRhoSc} and \Append{sDeriveStatCb}, we can compute for \CDD{} the correlation coefficient of the \SNR{} ($\RhoScCdd$) and of $C_b$ ($\RhoRbCdd$) and sum of those respectively ($\SumScCdd$, $\SumRbCdd$).
Although we can compute all these for the general \PDP{} ($\alphaim$), we assume for simplicity that \PDP{} in each antenna is the same, \ie{} $\alphaim = \alpha_{j,m}$ for $i \neq j$.
However, this assumption is very feasible because Tx antennas are not separated so much.
When we let $\Cov(H_\NOne,H_\NTwo)$ denote the covariance of \SISO{}, the covariance in \eqref{eCovCddHOri} reduces to
    \begin{equation}
        \label{eCovCddH}
             \Cov(\HnOneCdd, \HnTwoCdd) = \Cov(H_\NOne,H_\NTwo) \sum_{i=1}^{\NTx} \frac{e^{ -j \tfrac{2\pi}{\Nsc} D_i (\NTwo - \NOne)}} {\NTx}.
    \end{equation}
Noting that $\HnCdd$ follows the same distribution as that of $H_n$ and that $\GammanCdd = P |\HnCdd|^2 / \sigma_w^2$, we can easily have for the correlation coefficient between $\GammanOneCdd$ and $\GammanTwoCdd$ from \eqref{eRhoSnrScDerive} and \eqref{eCovCddH} as
    \begin{equation}
        \label{eRhoCddSnrScDerive}
            \RhoScCdd(|\Delta_n|)= \RhoSc(|\Delta_n|) \left| \frac{\sum_{i=1}^{\NTx} e^{ -j \tfrac{2\pi}{\Nsc} D_i (\Delta_n)}} {\NTx} \right|^2.
    \end{equation}
This shows that the correlation coefficient of \CDD{} is the correlation coefficient of \SISO{} ($\RhoSc(|\Delta_n|)$) multiplied by a weight function.
This weight function consists of sinusoidal functions each with period $\frac{\Nsc}{D_i}$, while $\RhoSc$ of \SISO{} has a period of $\Nsc$.
Thus, $\RhoScCdd$ is periodic with a period $\Nsc$.
We can easily verify that the magnitude of the weight function is less than or equal to 1.
We find for every $\Delta_n$ that $\RhoScCdd$ has a value between zero and $\RhoSc$ depending on the sinusoidal weight with a shorter period, which indicates $\RhoScCdd$ is more fluctuating than $\RhoSc$ with respect to $\Delta_n$.
That is, a channel of \CDD{} is more fluctuating than that of \SISO{}.

Using $\RhoScCdd$ in \eqref{eRhoCddSnrScDerive}, we can compute $\SumScCdd$ from \eqref{eSumSc}.
Once we compute $\SumScCdd(r,|\Delta_b|,\Srb)$, we can compute $\RhoRbCdd(|\Delta_b|)$ from \eqref{eRhoCapRb} and $\SumRbCdd(\Srb)$ from \eqref{eSumRb}.
From these and \eqref{eDPerUserOsApprExMaxCb} and \eqref{eDPerUserGaApprExMaxCb}, we can find $\DPerUserVec$ that maximizes $\ExMaxCb$ by exhaustive search.

\subsubsection{Role of cyclic delay on frequency selectivity}
\label{sRoleOfCd}
We mentioned that $\SumScInv$ represents frequency selectivity in \Section{sSumScInv} and also the effective number of paths in a channel or independent subcarriers in \Section{sSumScInv}.
For the better understanding about the role of cyclic delay ($D_i$) in $\SumScInvCdd$, let us consider a simple example.
Suppose that we have two transmit antennas ($\NTx=2$) and that the channel in each antenna has $L$-path uniform \PDP{}, \ie{} $\alpha_{1,m}=\alpha_{2,m}= \sqrt{1/L}$, for $1 \leq m \leq L$.
Cyclic delay is denoted by $\Dvec=[0,D]^T$ (\ie{} $D_2=D$).
We can easily verify in \eqref{eSumScInv} that the effective number of paths is $\SumScInv = L$ for each \SISO{} channel.
Suppose in \eqref{eCddChannel} that path-$m$ in a channel of Tx antenna-$1$ is overlapped with path-$(m-D)$ in a channel of Tx antenna-$2$.
Then, the average gain of \CDD{} in path-$m$ is $\AlphamCdd = \sqrt{(\alpha_{1,m}^2 + \alpha_{2,m-D}^2 )/2}$ since two channels are independent.

When $ D < L$, two \PDP{}s are overlapped for $D+1 \leq m \leq L$, but they are not in other range of $m$.
Following the way mentioned above, we have \PDP{} of \CDD{} as
    \begin{equation}
        \label{eCddPdp}
            \AlphamCdd =
                \begin{cases}
                    \frac{1}{\sqrt{2L}} \; , \hspace{0.5 cm} m \in [1,D] \textrm{ or } m \in [L+1,L+D] \\
                    \frac{1}{\sqrt{L}} \; , \hspace{0.5 cm} m \in [D+1,L]
                \end{cases}.
    \end{equation}
When $D >= L$, two \PDP{}s are not overlapped, and $\AlphamCdd = \sqrt{1 / 2L}$ for $m \in [1,L]$ and $m \in [D+1,D+L]$.
Then, the effective number of paths of \CDD{} is given from \eqref{eSumScInv} by
    \begin{equation}
        \label{eEffDivCdd}
            \SumScInvCdd =
                \begin{cases}
                    \frac{2 L^2}{2L - D}, \hspace{0.5cm} D < L \\
                    2L, \hspace{1cm} D \geq L
                \end{cases}.
    \end{equation}
From \eqref{eEffDivCdd}, we note that $\SumScInvCdd$ increases with $D$ for $D<L$, which indicates that the effective number of paths increases.
This agrees well with the fact that the number of paths in \CDD{} channel increases with $D$ for $D < L$.
However, $\SumScInvCdd$ does not increase any more for $D \geq L$.
This also agrees well since the number of paths in \CDD{} channel is always $2L$ in this range of $D$.
We can verify this situation even in more general case of a channel with not necessarily uniform \PDP.
Suppose just that $\alpha_{1,m} = \alpha_{2,m} = \alpha_m$.
Following the same way mentioned above to calculate \PDP{} of \CDD{} channel, we have
    \begin{equation}
        \label{eGenPdpCdd}
            \AlphamCdd =
                \begin{cases}
                    \hspace{0.6 cm} \frac{\alpham}{\sqrt{2}} \; , \hspace{1.1 cm} m \in [1,D] \textrm{ or } m \in [L+1,L+D] \\
                    \vspace{-0.7 cm}  \\
                    \frac{\sqrt{\alpham^2 + \alpha_{m-D}^2}}{\sqrt{2}} \; , \hspace{0.5 cm} m \in [D+1,L].
                \end{cases}
    \end{equation}
Then, the effective number of paths of \CDD{} is given from \eqref{eSumScInv} by
    \begin{equation}
        \label{eEffDivGenCdd}
            \SumScInvCdd =
                \begin{cases}
                    \frac{2}{\sum_{m=1}^{L}\alpha_m^4 + \sum_{m=D+1}^{L} \alpha_m^2 \alpha_{m-D}^2} \; , \hspace{0.5 cm} D < L \\
                    \vspace{-0.8 cm}  \\
                    \frac{2}{\sum_{m=1}^{L} \alpha_m^4} \; , \hspace{0.5 cm} D \geq L.
                \end{cases}
    \end{equation}
For $D<L$, we note that the first sum in the denominator is not affected by $D$.
We find that the number of product terms in the second sum of the denominator is $L-D$.
Thus, an increase of $D$ reduces the number of product terms, which leads to a decrease of the denominator and an increase of the effective number of paths.
This indicates that cyclic delay ($D$) increases the effective number of paths, which leads to an increase of the effective number of subcarriers or frequency selectivity.
We can also see that there is no more increase in $\SumScInvCdd$ for $D \geq L$.

\subsection{Determination of cyclic delay from $\TauRms$}
\label{sDetCdFromTauRms}
In \eqref{eDPerUserGaApprExMaxCb}, we need to maximize $\sqrt{\SumScCdd(1,0,\Srb) \ln \tfrac{1}{\SumRbCdd(\Srb)}}$ since $E_1$ and $V_1$ are constant with respect to $\Dvec$.
Considering \eqref{eSumScSumRb}, we need to maximize $\sqrt{\SumScCdd(1,0,\Srb) \ln \tfrac{\SumScCdd(1,0,\Srb)}{\SumScCdd(1,0,\Nsc)}}$.
In \Section{sApprMaxCb}, we found that the channel should be as flat as possible inside a block and as independent as possible between blocks to maximize multiuser diversity.
Coherence bandwidth is regarded as the bandwidth where correlation between any two frequency component is enough large or more specifically larger than or equal to a certain large threshold \cite{bZhangTcom07}.
In this section, we take the coherence bandwidth as the criteria for the flatness inside a block.
That is, we take that a channel is enough flat inside a block if block size is less than or equal to the coherence bandwidth.
This also implies that it is enough for $\SumScCdd(1,0,\Srb)$ to be larger than or equal to a certain threshold.
Under this assumption, we need to maximize $\SumScInvCdd$ from the equation mentioned above.

\begin{figure}
        \centering
            \includegraphics[width=8.0cm]{\DetCdFigDir/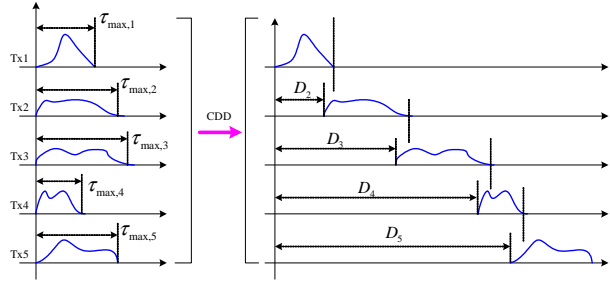}
            \caption{Example of power delay profile (\PDP) in a cyclic delay diversity (\CDD) channel. We note that \PDP{} for Tx-$i$ and Tx-$(i+1)$ does not overlap when $D_{i+1} > D_i + \TauMaxi$, so that frequency selectivity does not increase any more.}
            \label{fExPdpCdd}
    \end{figure}

In \eqref{eEffDivCdd} and \eqref{eEffDivGenCdd}, we note that $\SumScInvCdd$ does not increase when cyclic delay is larger than the number of paths.
More generally in \Fig{fExPdpCdd}, we cannot obtain any more gain in $\SumScInvCdd$ when any two \PDP{}s are not overlapped any more.
Therefore, we need an additional constraint that \PDP{} for Tx antenna-$i$ with cyclic delay $D_i$ should be overlapped with \PDP{} for Tx antenna-$(i+1)$ with cyclic delay $D_{i+1}$ as in \Fig{fExPdpCdd}.
From the above, the problem we focus on is
    \begin{equation}
        \label{eDetCdFromTauRms}
            \max_{\Dvec} \SumScInvCdd \hspace{0.1 cm} \textrm{ s.t. }
            \begin{cases}
                { \CohBwCdd \geq \Srb} \\
                {D_{i+1} \leq D_i + \TauMaxi, (1 \leq i < \NTx)},
            \end{cases}
    \end{equation}
where $\TauMaxi$ denotes the maximum delay spread in Tx antenna-$i$.
As in many applications of \CDD{} \cite{bBauchTwc06,bLiangTwc08}, we consider the case that $D_i = (i-1) D$.
Then, we find that $\SumScInvCdd$ increases with $D$ in \eqref{eEffDivCdd} and \eqref{eEffDivGenCdd}.
Since frequency selectivity increases with $D$, the coherence bandwidth decreases with $D$.
Let $\DnumBc$ denote a maximum cyclic delay to meet $\CohBwCdd \geq \Srb$.
Let $\DnumMax = \min_{1 \leq i < \NTx} \TauMaxi$.
Then, we note from \eqref{eDetCdFromTauRms} that cyclic delay which maximizes $\ExMaxCb$ is the maximum of $D$ while meeting two constraints of $D \leq \DnumBc$ and $D \leq \DnumMax$.
That is, we can reduce \eqref{eDetCdFromTauRms} to
    \begin{equation}
        \label{eDetCdFromTauRmsFinal}
            \DPerUser = \min \{ \DnumBc, \DnumMax \}.
    \end{equation}

\subsubsection{Coherence bandwidth of \CDD{} channel}
\label{sCohBw}
A root mean square (\RMS) delay spread following the notations in \eqref{eChannelForFreqDiv} is defined as \cite{bRappaportPh02}
    \begin{equation}
        \label{eTauRms}
                \TauRms = \sqrt{\sum_{m=1}^L (m-1)^2 \alpha_m^2 - \left( \sum_{m=1}^L (m-1) \alpha_m^2 \right)^2}.
    \end{equation}
This is widely used in characterizing frequency selectivity of a channel \cite{bRappaportPh02,bLeeMgh89,bZhangTcom07}.
When frequency selectivity increases (for example, the number of paths increases in a channel), $\TauRms$ increases in \eqref{eTauRms}.

Noting that the sum of a power delay profile (\PDP{}) is normalized to $1$ in \eqref{eChannelForFreqDiv}, we can regard a delay spread (or excess delay) $\tau$ in Tx antenna-$i$ as a random variable with a probability density function (\PDF{}) of $f_i(\tau) = \sum_{m=1}^{L_i} \alpha_{i,m}^2 \delta(\tau - m + 1)$.
Let $\mu_i = \Ex[\tau]$ and $\TauRmsi = \sqrt{\Var[\tau]}$ denote the average and \RMS{} delay spread in Tx antenna-$i$.
We note that this $\TauRmsi$ exactly matches with \eqref{eTauRms}.

We mentioned in \Section{sRoleOfCd} that \PDP{} in \CDD{} channel is the average of \PDP{} in each Tx antenna channel delayed by a cyclic delay.
This is noted in \Fig{fExPdpCdd} as well.
Considering this property, we have a \PDF{} for $\tau$ of \CDD{} channel using its \PDP{} as
    \begin{equation}
        \label{eFcdd}
                \fcdd(\tau) = \frac{1}{\NTx} \sum_{i=1}^{\NTx} f_i(\tau - D_i).
    \end{equation}
Then we can easily have the average delay spread as
    \begin{equation}
        \label{eMucdd}
                \mucdd = \int_{0}^{\infty} \tau \fcdd ( \tau ) \; d \tau = \frac{1}{\NTx} \sum_{i=1}^{\NTx} (\mu_i + D_i).
    \end{equation}
Noting that $\Var[\tau] = \Ex[\tau^2] - (\Ex[\tau])^2$, we also have for the \RMS{} delay spread as
    \begin{equation}
        \label{eTaucdd}
                \TauRmsCdd = \sqrt{\sum_{i=1}^{\NTx} \tfrac{ \TauRmsi^2 + (\mu_i + D_i)^2 }{\NTx}
                     - \left( \sum_{i=1}^{\NTx} \tfrac{\mu_i + D_i}{\NTx} \right)^2}.
    \end{equation}
When $D_i = (i-1)D$ as in \cite{bBauchTwc06,bLiangTwc08}, we can reduce \eqref{eTaucdd} to
\begin{equation}
    \label{eTaucdd2}
            \TauRmsCdd = \sqrt{aD^2 + bD + c + \TauRmsBar^{\; 2}},
\end{equation}
where $\TauRmsBar = \sqrt{\sum_{i=1}^{\NTx} \TauRmsiSq / \NTx}$ and other constants are defined as
    \begin{equation}
        \label{eConstSum}
        \begin{array}[c]{l|l}
            \hline
                a \triangleq \tfrac{1}{12} (\NTx^2 - 1)  &   \muone \triangleq \tfrac{1}{\NTx} \sum_{i=1}^{\NTx} \mu_i    \\
            \hline
                b \triangleq 2 \muw - \muone (\NTx + 1) &   \muw \triangleq \tfrac{1}{\NTx} \sum_{i=1}^{\NTx} i \mu_i    \\
            \hline
                c \triangleq \mutwo - (\muone)^2    &   \mutwo \triangleq \tfrac{1}{\NTx} \sum_{i=1}^{\NTx} \mu_i^2  \\
            \hline
        \end{array} \hspace{0.1cm}.
    \end{equation}

Since the channel coherence bandwidth can be represented as the inverse of the $\RMS$ delay spread \cite{bRappaportPh02,bZhangTcom07,bLeeMgh89}, the coherence bandwidth of \CDD{} channel is given by
\begin{equation}
    \label{eBcCdd}
            \CohBwCdd \simeq \frac{1}{K \TauRmsCdd} = \frac{1}{K \sqrt{ aD^2 + bD + c + \TauRmsBar^{\; 2}}}
\end{equation}
where $K$ is a constant to determine the coherence bandwidth, which is related to the minimum correlation coefficient of the \SNR{} between two frequency components within the coherence bandwidth.

\subsubsection{Relation between the maximum delay spread $(\TauMaxi)$ and the \RMS{} delay spread $(\TauRmsi)$}
\label{sTauMax}
For the delay spread $\tau$ in the channel of Tx antenna $i$ with mean $\mu_i$ and variance $\TauRmsiSq$, we have from the Chebyshev inequality \cite{bGarciaAw94}
\begin{equation}
    \label{eChebyshev}
            \int_{| \tau - \mu_i | \leq \epsilon} f_i(\tau) d \tau = Pr\{ | \tau - \mu_i | \leq \epsilon \} \geq 1 - \frac{\TauRmsiSq}{\epsilon^2} \triangleq \kappa.
\end{equation}
This inequality indicates that the ratio of the total received power to the transmitted power is equal to or greater than $\kappa$ when $| \tau - \mu_i | \leq \epsilon$, \ie{} $\mu_i - \epsilon \leq \tau \leq \mu_i + \epsilon$.
For example, $\kappa=0.9$ means that the received power is over $90\%$ of the transmitted power in that range of $\tau$.
Then, we have from \eqref{eChebyshev}
\begin{equation}
    \label{eKappa}
            \epsilon = \frac{\TauRmsi}{\sqrt{1 - \kappa}}.
\end{equation}
If we let the maximum delay spread $\TauMaxi$ be the length of the delay spread where the power ratio is equal to or larger than $\kappa$ and we let $\TauMaxi$ be an integer for the later use for cyclic delay, $\TauMaxi$ is given by
\begin{equation}
    \label{eTauMax}
            \TauMaxi = [\mu_i + \epsilon] - [\mu_i - \epsilon] + 1,
\end{equation}
where $[x]$ indicates the maximum integer that is not greater than $x$.
Since $\TauMaxi < 2 \epsilon + 2$ in \eqref{eTauMax} and it is an integer, we have from \eqref{eKappa}
\begin{equation}
    \label{eTauMax2}
            \TauMaxi =  \lceil 2 \epsilon \rceil + 1 =  \left \lceil \frac{2 \TauRmsi}{\sqrt{1 - \kappa}} \right \rceil + 1.
\end{equation}

\subsubsection{Determine $\DPerUser$}
\label{sDetDPerUser}
From \eqref{eBcCdd}, the maximum cyclic delay $\DnumBc$ in \eqref{eDetCdFromTauRmsFinal} to meet $\CohBwCdd \geq \Srb$ is given by
    \begin{equation}
        \label{eDnumBc}
                \DnumBc = \Big[ \tfrac{1}{\sqrt{a}} \; \sqrt{ \left( \tfrac{1}{K^2 \Srb^2} - \TauRmsBar^2 + \tfrac{b^2- 4ac}{4a} \right)_+} - \tfrac{b}{2a} \Big],
    \end{equation}
where $(x)_{\SubPlus}$ denotes $\max (0,x)$.
From \eqref{eTauMax2}, $\DnumMax$ in \eqref{eDetCdFromTauRmsFinal} is given by
    \begin{equation}
        \label{eDnumMax}
                \DnumMax = \min_{1 \leq i < \NTx} \left \lceil \tfrac{2 \TauRmsi}{\sqrt{1 - \kappa}} \right \rceil + 1.
    \end{equation}
Then, we have the per-user optimal cyclic delay $\DPerUser$ in \eqref{eDetCdFromTauRmsFinal} as the minimum of $\DnumBc$ in \eqref{eDnumBc} and $\DnumMax$ from \eqref{eDnumMax}.

To have an idea about the relation between $\DPerUser$ and the \RMS{} delay spread and block size $\Srb$, let us consider a simple and practical case.
Suppose that channels in all Tx antennas have the same average delay spread and the same \RMS{} delay spread, \ie{} $\TauRmsi = \TauRmsj = \TauRms$ and $\mu_i = \mu_j$ $(1 \leq i,j \leq \NTx)$.
We note that we do not put any other constraint on \PDP's of channels.
After some manipulation, we have for per-user optimal cyclic delay as
    \begin{equation}
        \label{eDPerUserCddByTauRms}
                \DPerUser = \min \Big \{ \sqrt{ \tfrac{12}{{\NTx^2 - 1}}  \left( \tfrac{1}{K^2 \Srb^2} - \TauRmsSq \right)_+} \; , \left \lceil \tfrac{2 \TauRms}{\sqrt{1 - \kappa}} \right \rceil + 1 \Big \}
    \end{equation}
We note in \eqref{eDPerUserCddByTauRms} that $\DPerUser$ increases with $\TauRms$ for the small $\RMS$ delay spread because the second term is dominant.
When $\TauRms$ is large, the first term is dominant and $\DPerUser$ decreases with $\TauRms$.
For example, in flat fading channel, $\DPerUser=1$ because $\TauRms=0$, which agrees with the idea that there is no more gain in effective diversity ($\SumScInvCdd$) for larger cyclic delay than $1$.
We also note that $\DPerUser$ should become smaller as $\Srb$ grows larger.
This agrees well with the idea that a large block size requires a large coherence bandwidth and thus small cyclic delay.

When frequency selectivity in a given channel is already large enough, $\TauRmsSq$ in \eqref{eDPerUserCddByTauRms} makes the first term zero, and $\DPerUser$ reduces to zero.
This indicates that \CDD{} does not give any benefit for $\ExMaxCb$ in this channel.
From this, we note that there may exist an optimal threshold of $\TauRms$ whereby we decide whether to employ \CDD{} or not to enhance multiuser diversity, which is left as a future work.
We can also say that this threshold decreases with $\Srb$.

\section{Numerical Results}
\label{sNumericalResults4FreqDiv}
To obtain numerical results, we consider $\Nsc=1024$ for the \FFT{} size and exponential \PDP{} for each channel of Tx antenna as following.
    \begin{equation}
        \label{eExpPdp}
            \alpha_m = \tfrac{e^{- \frac{m}{\tau_o}} }{ \sqrt{\sum_{m=1}^{L} e^{- \frac{2m}{\tau_o}}}}.
    \end{equation}
Various \RMS{} delay spreads are obtained by changing $\tau_o$ in $\alpha_m$.
We consider that the number of paths $L$ is less than or equal to $64$ depending on the \RMS{} delay spread.
For each obtained channel, we compute all functions in \Section{sDefineUsefulFncForFreqSel} for numerical evaluation of maximum of the block average throughput ($\ExMaxCb$).
For comparison purpose, we show Monte-Carlo simulation results for maximum of the block average throughput ($\ExMaxCb$) and the sum rate ($\Rsum$) using proportional fair scheduling described in \Section{sPropFairScheduling}.
Regarding \CDD{}, we use $\NTx=2$ to better characterize the role of cyclic delay.

\subsection{Frequency selectivity, intra-block sum correlation and the effective number of blocks}
\begin{figure}
        \centering
            \hspace{-0.0cm}\includegraphics[width=\FigWidth,height=\FigHeight]{\DetCdFigDir/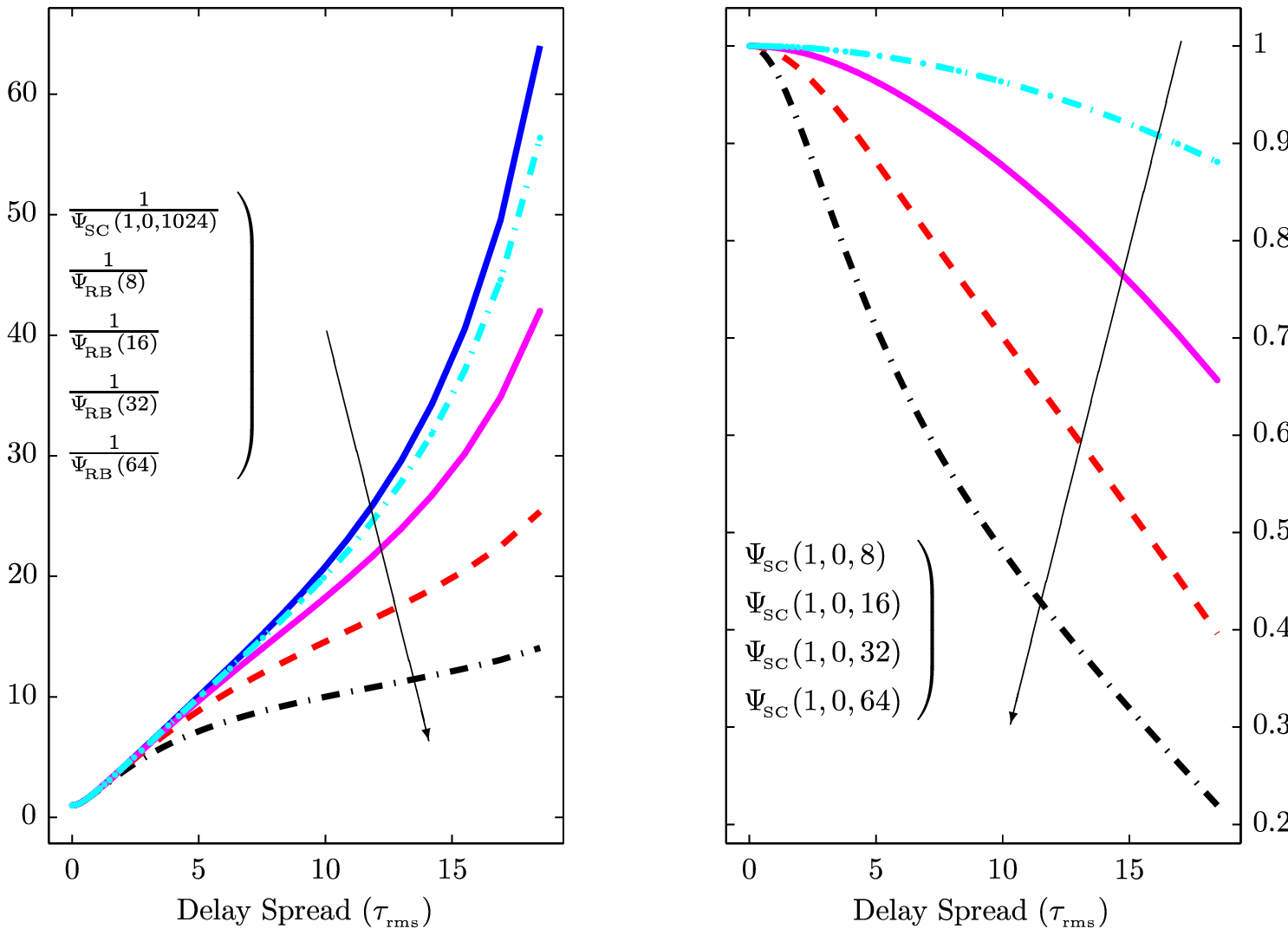}\\
            \vspace{-0.4cm}
            \begin{tabular}[c]{cc}
                {\hspace{-0.0cm} \scriptsize (a)} & {\hspace{4.5cm} \scriptsize (b)}
            \end{tabular}
            \vspace{-0.4cm}
            \caption{Effect of frequency selectivity ($\TauRms$) on the effective number of subcarriers ($\SumScInv$), the effective number of blocks ($\SumRbInv$), and the intra-block sum correlation ($\SumSc(1,0,\Srb)$).}
            \label{fSumScSumRb}
    \end{figure}

When $\tau_o$ in \eqref{eExpPdp} increases, both the \RMS{} delay spread and the number of valid paths increase.
Thus, frequency selectivity measure $\SumScInv$, also known as the effective number of paths, increases with the \RMS{} delay spread in \Fig{fSumScSumRb}(a).
This also explains an increase of the effective number of blocks\footnote{We note that the effective number of blocks is the inverse of inter-block sum correlation as discussed in \Section{sRhoRb}.} $\SumRbInv$ for each block size in \Fig{fSumScSumRb}(a).
Meanwhile, correlation between subcarriers decreases and thus the intra-block sum correlation decreases with the \RMS{} delay spread in \Fig{fSumScSumRb}(b).
Since the effective number of blocks increases but the intra-block sum correlation decreases with the \RMS{} delay spread in this figure, we can verify the trade-off between them in \eqref{eOsApprExMaxCb} and \eqref{eGaApprExMaxCb}.

\begin{figure}
        \centering
            \hspace{-0.0cm}\includegraphics[width=\FigWidth,height=\FigHeight]{\DetCdFigDir/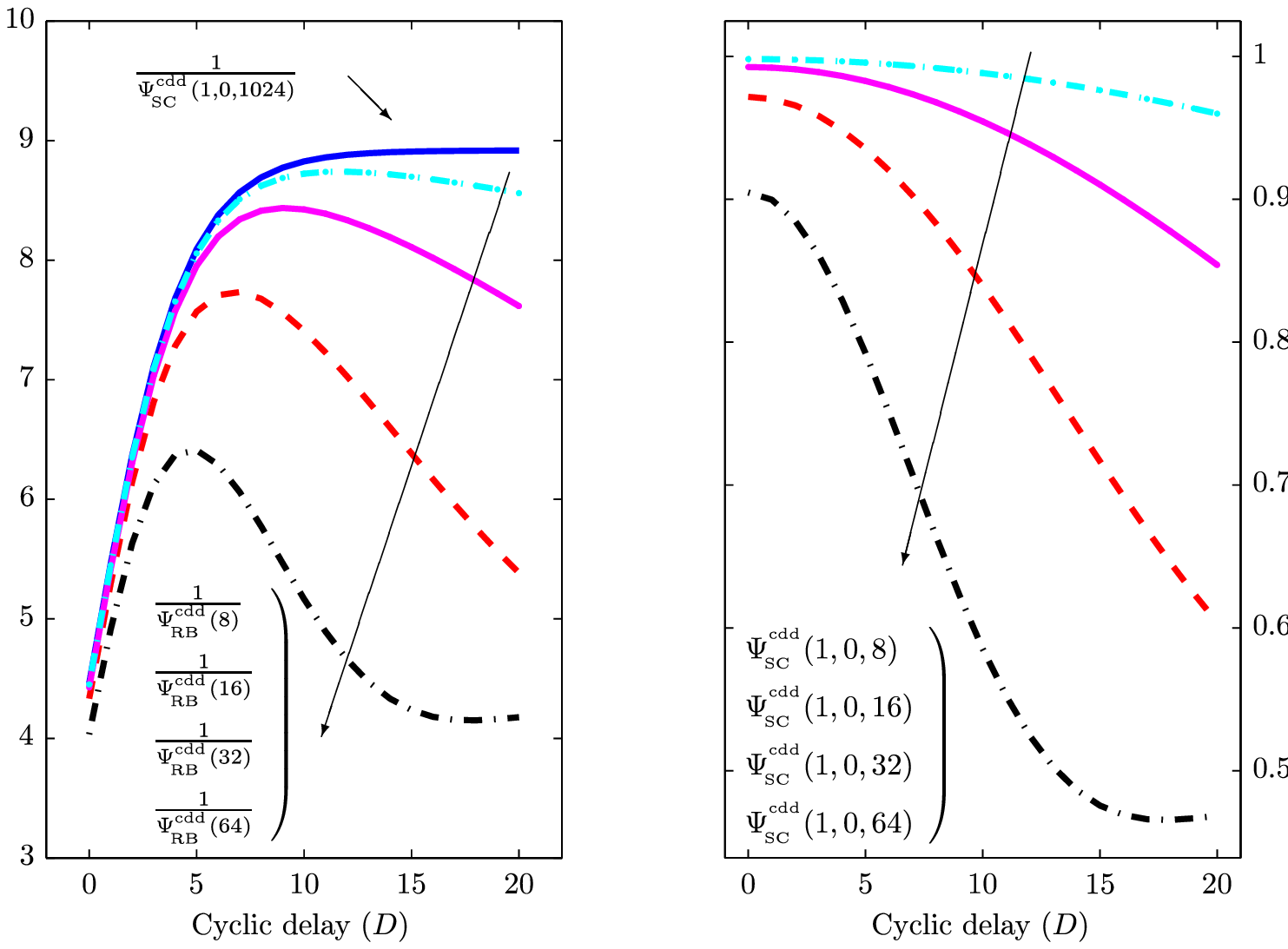}\\
            \vspace{-0.4cm}
            \begin{tabular}[c]{cc}
                {\hspace{-0.0cm} \scriptsize (a)} & {\hspace{4.5cm} \scriptsize (b)}
            \end{tabular}
            \vspace{-0.4cm}
            \caption{Effect of cyclic delay ($D$) on the effective number of subcarriers ($\SumScInv$), the effective number of blocks ($\SumRbInv$), and the intra-block sum correlation ($\SumSc(1,0,\Srb)$).}
            \label{fSumScCddSumRbCdd}
    \end{figure}

As discussed in \Section{sAddFreqSel}, frequency selectivity increases with cyclic delay in \CDD{}.
We can verify this in \Fig{fSumScCddSumRbCdd}(a).
In a different way from \Fig{fSumScSumRb}, frequency selectivity saturates to two times of the value for \SISO{} (\ie{} $D=0$).
This confirms the discussion in \Section{sRoleOfCd} that the number of paths does not increase when cyclic delay is larger than the number of paths in a given channel.
As cyclic delay increases, the sinusoidal components in \eqref{eRhoCddSnrScDerive} cause more local peaks in correlation because the period $\frac{\Nsc}{D}$ decreases.
This makes block correlation larger and the effective number of blocks does not increase monotonically with cyclic delay in \Fig{fSumScCddSumRbCdd}(a).
Meanwhile, we note that the intra-block sum correlation always decreases with cyclic delay in \Fig{fSumScCddSumRbCdd}(b).
We can find the trade-off between $\SumRbInvCdd$ and $\SumScCdd(1,0,\Srb)$ with respect to cyclic delay.
However, for the larger cyclic delay than that which gives the peak of $\SumRbInv$, both of effective number of blocks and the intra-block sum correlation decrease.
Thus, we don't have to consider these cyclic delays for evaluation of \eqref{eDPerUserOsApprExMaxCb} and \eqref{eDPerUserGaApprExMaxCb}, which much saves the load of exhaustive search.

\subsection{Optimality of frequency selectivity on multiuser diversity and optimal addition of frequency selectivity}
\begin{figure}
        \centering
            \includegraphics[width=\FigWidth,height=\FigHeight]{\DetCdFigDir/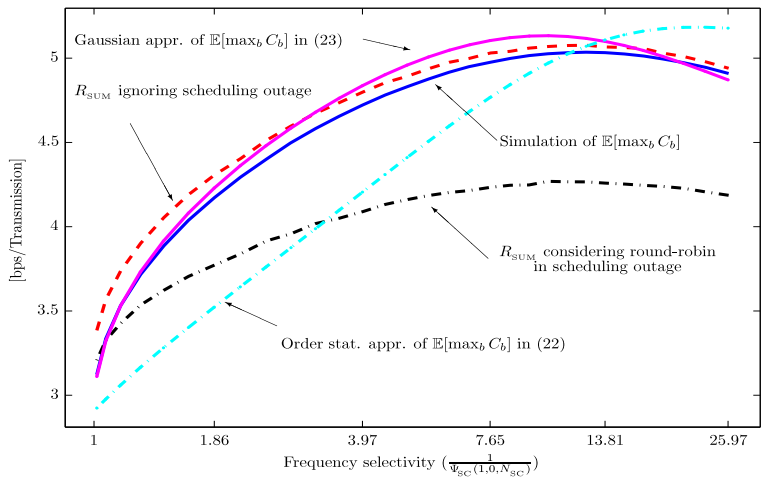}
            \caption{Effect of frequency selectivity ($\SumScInv$) on multiuser diversity (\ie{} maximum of the block average throughput of an arbitrary user ($\ExMaxCb$) and the sum rate ($\Rsum$)). Two approximations for $\ExMaxCb$ in \eqref{eOsApprExMaxCb} and \eqref{eGaApprExMaxCb} are compared as well. `RR' denotes round robin scheduling. In the case of without RR, blocks are ignored when a scheduling outage happens. ($\Nrb$=32 blocks, $K$=32 users)}
            \label{fExMaxCb}
    \end{figure}

In \Fig{fExMaxCb}, we first note that Gaussian approximation of $\ExMaxCb$ in \eqref{eGaApprExMaxCb} better matches with the simulation than order statistic approximation in \eqref{eOsApprExMaxCb}.
Further, when we do not consider a round robin scheduling for the scheduling outage (\ie{} no user reports for a block), Gaussian approximation and the simulation result of $\ExMaxCb$ are well matched with the simulation result of the sum rate.
This can justify the Gaussian approximation of the block average throughput.
We note that there exists optimal frequency selectivity that maximizes the sum rate.
Since maximizing $\ExMaxCb$ is related to the per-user optimality, we also note that per-user optimality is good for the approximation of the sum rate optimality.
When we use a round-robin scheduling for blocks in scheduling outage, an arbitrary user is selected for those blocks.
This causes the sum rate to decrease compared to other cases.
However, optimal frequency selectivity is not changed much.
We also find that the sum rate in a limited fluctuating channel with small frequency selectivity is very small.
This implies that addition of frequency selectivity would enhance the sum rate as in \CDD.

\begin{figure}
        \centering
            \includegraphics[width=\FigWidth,height=\FigHeight]{\DetCdFigDir/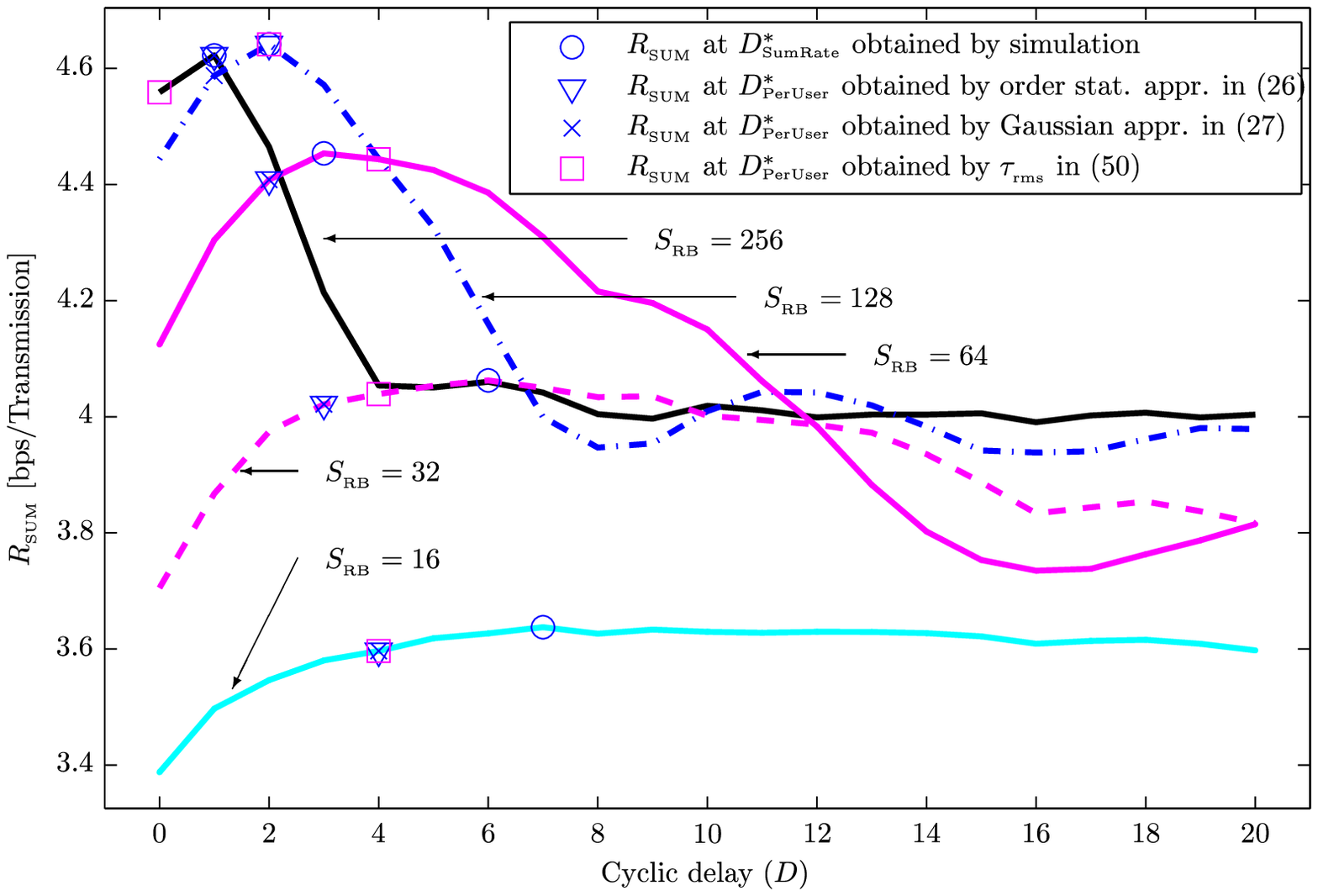}
            \caption{Effect of cyclic delay ($D$) on the sum rate. In each curve for the different block size, $\DSumRate$ and $\DPerUser$ are marked. ($\SumScInv=1.6246$ of original channel, $K=32$ users)}
            \label{fExMaxCbCdd}
    \end{figure}

\begin{figure}
        \centering
            \includegraphics[width=\FigWidth,height=\FigHeight]{\DetCdFigDir/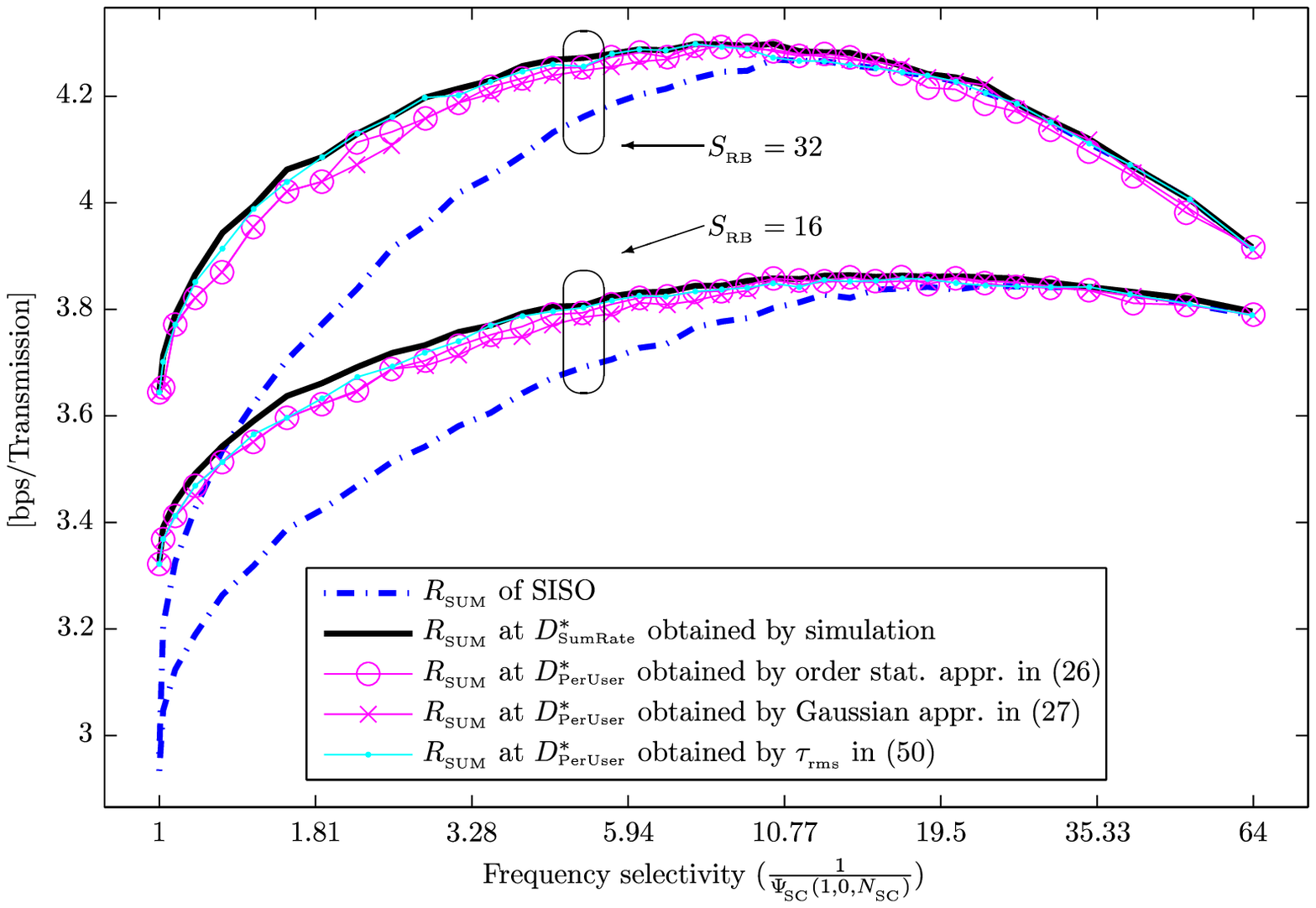}
            \caption{Sum rate comparison for \SISO{}, \CDD{} with per-user optimal cyclic delay $\DPerUser$ and \CDD{} with sum-rate optimal cyclic delay $\DSumRate$ as a function of frequency selectivity ($\SumScInv$). Two approximations in \eqref{eDPerUserOsApprExMaxCb} and \eqref{eDPerUserGaApprExMaxCb} are used for $\DPerUser$. ($K=32$ users)}
            \label{fSumRateGainCddvsFreqSel}
    \end{figure}

\Fig{fExMaxCbCdd} shows the sum rate change with cyclic delay when \CDD{} is used to increase frequency selectivity.
First, we find from simulation results that the sum rate gain by \CDD{} to \SISO{} (\ie{} $D=0$) is remarkable and that there exists optimal cyclic delay in the sense of maximum sum rate.
In the figure, we mark per-user optimal cyclic delays found by two approximations in \eqref{eDPerUserOsApprExMaxCb} and \eqref{eDPerUserGaApprExMaxCb} and the \RMS{} delay spread in \eqref{eDPerUserCddByTauRms}.
Although per-user optimality is not perfectly matched with sum-rate optimality, the sum rate by per-user optimal cyclic delay is very close to that by sum-rate optimal one.
This is also found in \Fig{fSumRateGainCddvsFreqSel}, which illustrates the sum rate of \CDD{} with $\DPerUser$ and $\DSumRate$ and the sum rate of a \SISO{} system.
We note that $\DPerUser$ achieves very close performance of $\DSumRate$.
We find that the gain of \CDD{} to \SISO{} system is remarkable especially in the range of small frequency selectivity, but small in a channel with large frequency selectivity.
This is because the achievable gain itself is small for a channel with frequency selectivity already close to optimal selectivity as shown in \Fig{fExMaxCb}.
This also shows the reason why all the schemes related random beamforming \cite{bViswanathTit02,bSharifTit05} are considered in a channel with slow fading at the time domain.

\begin{figure}
        \centering
            \includegraphics[width=\FigWidth,height=\FigHeight]{\DetCdFigDir/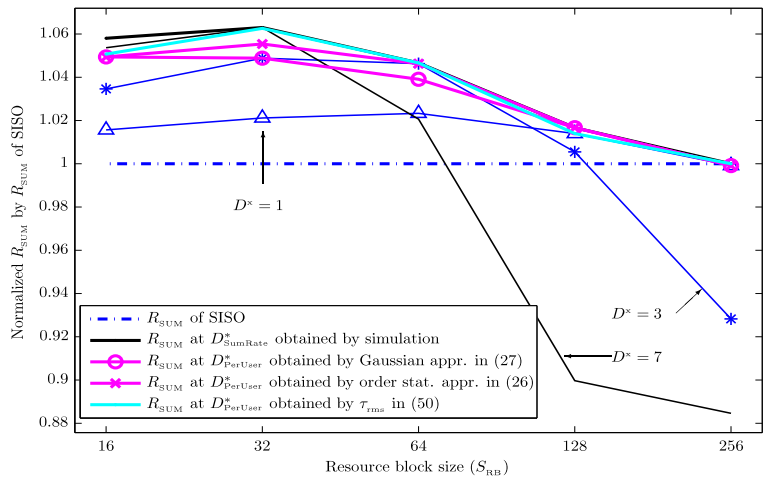}
            \caption{Sum rate gain of cyclic delay diversity compared to \SISO{} by per-user optimal cyclic delay $\DPerUser$ and sum-rate optimal cyclic delay $\DSumRate$ as a function of block size ($\Srb$). Two approximations for $\DPerUser$ in \eqref{eDPerUserOsApprExMaxCb} and \eqref{eDPerUserGaApprExMaxCb} and fixed cyclic delay scheme are compared as well. ($\SumScInv=1.6246$ of original channel, $K=32$ users)}
            \label{fSumRateGainCddvsSrb}
    \end{figure}

In \Fig{fSumRateGainCddvsSrb}, we compare the sum rate gain to \SISO{} for our $\DPerUser$ and arbitrarily fixed cyclic delay ($\Dfix$).
We find that $\DPerUser$ shows more stable and better performance than any fixed one in the whole range of block sizes.
In particular, misuse of cyclic delay leads to the smaller sum rate than that of \SISO.
This implies that adaptive cyclic delay based on our technique is better.
The case that fixed cyclic delay shows better performance in a specific $\Srb$ is corresponding to the case that fixed one happens to coincide with $\DSumRate$.

\subsection{Factors to affect optimal frequency selectivity}
\begin{figure}
        \begin{minipage}[b]{0.5\linewidth}
            \centering
            \includegraphics[width=8.0cm,height=5.0cm]{\DetCdFigDir/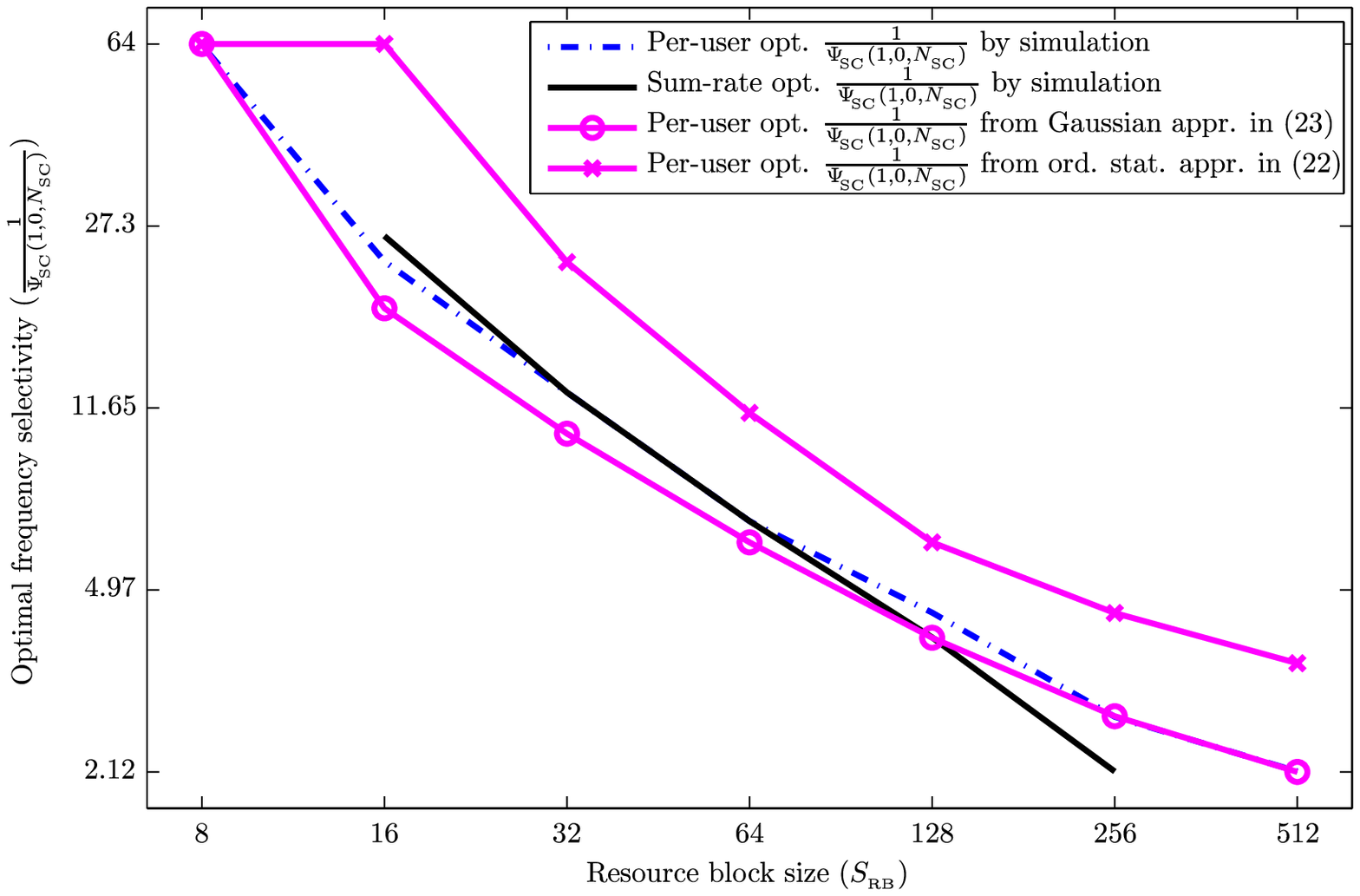}\\
            {\scriptsize (a)}
        \end{minipage}
        \begin{minipage}[b]{0.5\linewidth}
            \centering
            \includegraphics[width=8.0cm,height=5.0cm]{\DetCdFigDir/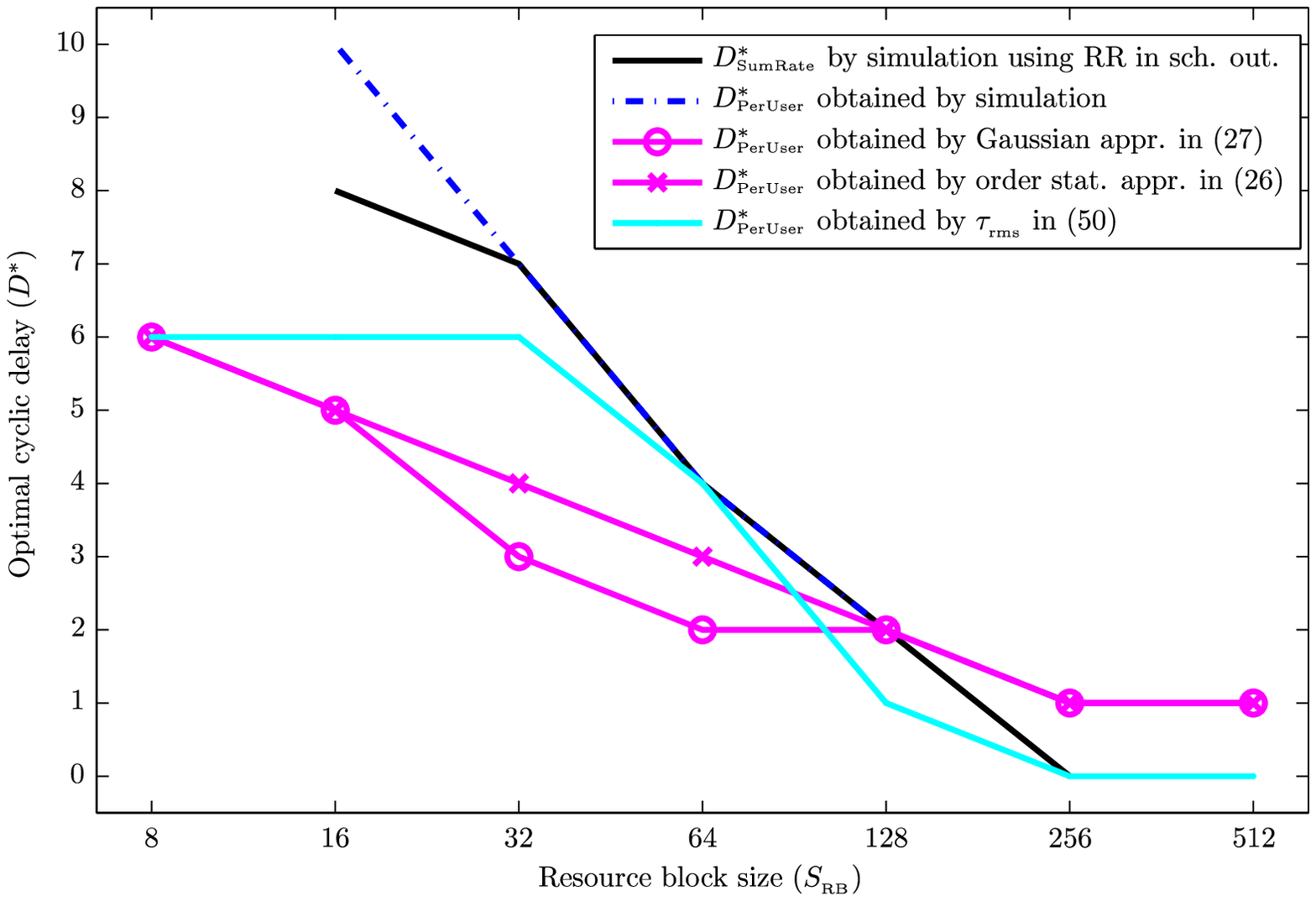}\\
            {\scriptsize (b)}
        \end{minipage}
            \caption{Effect of block size ($\Srb$) on optimal frequency selectivity ($\SumScInv$) (left) and on the optimal cyclic delay (right). Two approximations are compared to a simulation result. The simulated sum rate optimal one is compared as well. ($K$=32 users)}
            \label{fOptFreqSel}
    \end{figure}

We saw in \Fig{fSumScSumRb}(b) that the intra-block sum correlation $\SumSc(1,0,\Srb)$ in \eqref{eSumScOneZeroOri} decreases much in large block size for a small increase of frequency selectivity.
However, the effective number of blocks $\SumRbInv$ does not increase much in \Fig{fSumScSumRb}(a).
Thus, optimal frequency selectivity or cyclic delay that maximizes the trade-off in \eqref{eGaApprExMaxCb} and \eqref{eDPerUserGaApprExMaxCb} decreases with the block size, both of which are illustrated in \Fig{fOptFreqSel}(a) and \Fig{fOptFreqSel}(b), respectively.
We also find that per-user optimal frequency selectivity obtained by Gaussian approximation agrees well with that by simulation and with sum-rate optimal frequency selectivity except for $\Srb=256$ in \Fig{fOptFreqSel}(a).
Although cyclic delay calculated by approximation is not well matched with sum-rate optimal one, we stress again that the sum rate is close to optimal value as in \Fig{fSumRateGainCddvsFreqSel}.
When $\Srb=256$ and $K=32$ in the figure, there are 4 blocks.
Thus, about 8 users in the average sense contend for each block to be scheduled.
Thus, variance of a block becomes a more important factor and thus the large intra-block sum correlation is preferred to improve the sum rate.
This explains that frequency selectivity or cyclic delay for sum-rate optimality is smaller than that expected by the approximation in \Fig{fOptFreqSel}.

\begin{figure}
        \centering
            \includegraphics[width=\FigWidth,height=\FigHeight]{\DetCdFigDir/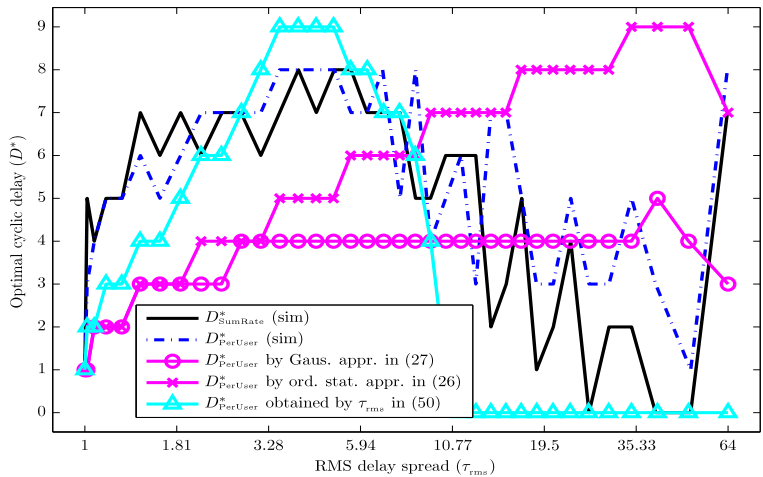}
            \caption{Comparison of per-user optimal cyclic delay ($\DPerUser$) and sum-rate optimal cyclic delay ($\DSumRate$) as a function of frequency selectivity ($\TauRms$). Two approximations for $\DPerUser$ in \eqref{eDPerUserOsApprExMaxCb} and \eqref{eDPerUserGaApprExMaxCb} are compared as well. ($\Nrb=32$ blocks,$\SumScInv=1.6246$ of original channel, $K=32$ users)}
            \label{fOptCd}
    \end{figure}

Frequency selectivity of a given channel is another factor to affect the optimal cyclic delay.
In \Fig{fOptCd}, we find that both of per-user optimal cyclic delay and sum-rate optimal cyclic delay increase with small frequency selectivity, but decrease with large frequency selectivity.
This indicates that an increase of diversity (\ie{} effective number of blocks, $\SumRbInv$) is dominant in a limited fluctuated channel.
However, making a variance large by keeping $\SumSc(1,0,\Srb)$ large is more important in a channel with large selectivity.
In a system employing a fixed cyclic delay without updating \PDP{}, we note in \Fig{fSumRateGainCddvsSrb} and \Fig{fOptCd} that large sum rate is achieved in rather small block size such as $\Srb \leq 64$ when we use $\Dfix= 3, 4,$ or $5$ suggested by $\DPerUser$.

\section{Conclusion}
\label{sConclusion4FreqDiv}
In this paper, we studied the effect of frequency selectivity on multiuser diversity.
We focused on analyzing maximum of the block average throughput of an arbitrary user by considering two approximations for that.
From these approximations, we found that there exists optimal frequency selectivity in the sense of maximizing multiuser diversity, and we verified this by a simulation as well.
We showed that the optimal channel is flat within a block and mutually independent between blocks.

Motivated by the fact that cyclic delay diversity (\CDD) increases a channel fluctuation, we considered to use \CDD{} in a channel with small frequency selectivity to enhance the sum rate of a system.
Based on the previous study of optimal frequency selectivity, we proposed two techniques to determine per-user optimal cyclic delay exploiting approximations we developed for multiuser diversity.
We investigated the role of cyclic delay to frequency selectivity as well.
We showed by simulation that the proposed techniques achieve better performance than a conventional fixed cyclic delay scheme and that the throughput is very close to the optimal sum rate possible with \CDD{}.

\appendices
\section{Derivation of $\RhoSc(|\Delta_n|)$}
\label{sDeriveRhoSc}
Let $\xvec = [H_\NOne,H_\NTwo]^T$ for $H_n$ in \eqref{eHn}.
Since we assume that $H_n$'s follow jointly Gaussian distribution, $\xvec$ follows $\CN(\zerovec,R_\xvec)$ where $R_\xvec$ denotes a covariance matrix and its elements are in \eqref{eCovH}.
Considering $\gamma_n = P |H_n|^2 / \sigma_w^2$ in \Section{sSystemModel4FreqDiv} and using $R_\xvec$, we have the general order correlation as \cite[2.14 in p.86]{bMillerAw74}
    \begin{equation}
        \label{eGenCorrSnr}
                 \Ex[\gamma_\NOne^\alpha \gamma_\NTwo^\beta]= \gammaBar^{\alpha + \beta} \alpha! \beta! \hspace{-0.2cm} \sum_{m=0}^{\min\{\alpha,\beta\}} \hspace{-0.2cm} \binom{\alpha}{m} \binom{\beta}{m} |\Cov(H_\NOne,H_\NTwo)|^{2m}
    \end{equation}
where $\gammaBar = \Ex[\gamma_n]$.
Then, covariance is given by
    \begin{equation}
        \label{eCovSnr}
                 \Cov(\gamma_\NOne,\gamma_\NTwo) = \Ex[\gamma_\NOne \gamma_\NTwo] - \gammaBar^2= \gammaBar^2 |\Cov(H_\NOne,H_\NTwo)|^{2}.
    \end{equation}
Noting that $\Cov(H_n,H_n)=1$ in \eqref{eCovH}, we have $\Var[\gamma_n]=\gammaBar^2$ in \eqref{eCovSnr}.
Using these results and following the definition of the correlation coefficient in \eqref{eRhoSnrSc}, we lead to \eqref{eRhoSnrScDerive}.

\section{Statistics of $C_b$}
\label{sDeriveStatCb}
Noting that $\gamma_n$ follows Gamma distribution and is identically distributed over $n$, we have without loss of generality
    \begin{equation}
        \label{eExpCb}
            \Ex[C_b]= \Ex[\log_2(1+\gamma_1)] = \frac{e^{\tfrac{\sigma_w^2}{P}} \; \Ei(1,\tfrac{\sigma_w^2}{P})}{\ln 2}
    \end{equation}
where $\Ei (a,x) = \int_{1}^{\infty} e^{- x t} t^{- a} dt$ \cite{bAbramowitz70} and the integral equality in \cite[4.337.2 in p.603]{bGradshteynAp00} is used as following.
    \begin{equation}
        \hspace{-0.0cm} \int_{0}^{\infty} e^{- x t} \ln (1 + y t) dt = \frac{1}{x} e^{\frac{x}{y}} \Ei \left(1, \frac{x}{y} \right).
    \end{equation}

Instead of using a slowly converging infinite series in computing $\Cov(\log_2(1+\gamma_\NOne),\log_2(1+\gamma_\NTwo))$ \cite{bMcKayTit08}, we use the delta method which is known as the Taylor series method \cite{bAssaliniTvt09}.
When we take the Taylor series expansion of $\log_2(1+\gamma_n)$ about $\Ex[\gamma_n]$, we have \cite{bAssaliniTvt09}
    \begin{equation}
        \label{eTaylor}
            \log_2(1+\gamma_n) = \log_2(1+\Ex[\gamma_n]) + \sum_{m=1}^\infty \frac{(-1)^{m-1}(\gamma_n - \Ex[\gamma_n])^m}{m (1 + \Ex[\gamma_n])^m \ln 2 }.
    \end{equation}
For the first order expansion of $\log_2(1+\gamma_n)$ in \eqref{eTaylor} (\ie{} $m=1$), we have from \eqref{eCb}
    \begin{equation}
        \label{eCbFoa}
                C_b \simeq \log_2(1 + \Ex[\gamma_n]) + \frac{1}{\Srb} \sum_{n = (b-1)\Srb}^{b\Srb} \frac{\gamma_n - \Ex[\gamma_n]}{(1+\Ex[\gamma_n])\ln 2}.
    \end{equation}
Using the bilinear property of covariance \cite{bCovUrl} and considering that covariance does not change by the addition of a constant and that $\Cov(\gamma_\NOne,\gamma_\NTwo)= \Var[\gamma_1] \RhoSc(|\Delta_n|)$ in \eqref{eRhoSnrSc}, covariance between $C_\BOne$ and $C_\BTwo$ is given by
    \begin{equation}
        \label{eCovCb1Cb2Foe}
                \Cov(C_\BOne, C_\BTwo) = \frac{\Var[\gamma_1]}{((1 + \Ex[\gamma_1]) \ln 2)^2} \; \SumSc(1,|\Delta_b|,\Srb).
    \end{equation}
From \eqref{eRhoCapRb} and the fact that $\Var[C_b] = \Cov(C_b,C_b)$, we have
    \begin{equation}
        \label{eVarCbFoe}
                \Var[C_b]= \frac{\Var[\gamma_1]}{\{ (1 + \Ex[\gamma_1]) \ln 2 \}^2} \SumSc(1,0,\Srb).
    \end{equation}
Thus, the correlation coefficient between $C_\BOne$ and $C_\BTwo$ is given by \eqref{eRhoCapRbFoa}.

For the second order expansion of $\log_2(1+\gamma_n)$ in \eqref{eTaylor} (\ie{} $m=2$), we have
    \begin{equation}
        \label{eLogSoe}
                \log_2 (1 + \gamma_n) = A_1 + A_2 \gamma_n + A_3 \gamma_n^2
    \end{equation}
where $A_1= \log_2(1+\Ex[\gamma_n]) - \tfrac{\Ex[\gamma_n]}{(1+\Ex[\gamma_n])\ln 2} - \tfrac{\Ex^2[\gamma_n]}{2 (1+\Ex[\gamma_n])^2 \ln 2} $, $A_2 = \tfrac{1}{(1+\Ex[\gamma_n])\ln 2} + \tfrac{\Ex[\gamma_n]}{(1+\Ex[\gamma_n])^2 \ln 2}$, and $A_3 = \tfrac{-1}{2 (1+\Ex[\gamma_n])^2 \ln 2}$.
From \eqref{eCb}, we have
    \begin{equation}
        \label{eCbSoe}
            C_b \simeq A_1 + A_2 \hspace{-0.3cm} \sum_{\substack{ n = 1 + \\(b-1)\Srb }}^{b \Srb} \hspace{-0.2cm} \frac{\gamma_n}{\Srb} + A_3 \hspace{-0.3cm} \sum_{\substack{ n = 1 + \\(b-1)\Srb }}^{b \Srb} \hspace{-0.2cm} \frac{\gamma_n^2}{\Srb}.
    \end{equation}
From \eqref{eGenCorrSnr} and \eqref{eRhoSnrScDerive}, we have in the same way as \eqref{eCovSnr}
\begin{eqnarray}
    \label{eCovGamma2}
            \hspace{-0.5cm} \Cov(\gamma_\NOne,\gamma_\NTwo^2) \hspace{-0.2cm} &=& \hspace{-0.2cm} 4 \Ex^3[\gamma_n] \RhoSc(|\Delta_n|),\\
    \label{eCovGamma3}
            \hspace{-0.5cm} \Cov(\gamma_\NOne^2,\gamma_\NTwo^2) \hspace{-0.2cm} &=& \hspace{-0.2cm} 4 \Ex^4[\gamma_n] \; \{4 \RhoSc(|\Delta_n|) + \RhoScSq(|\Delta_n|) \}. 
\end{eqnarray}
From \eqref{eCbSoe}, \eqref{eCovGamma2}, \eqref{eCovGamma3} and the bilinear property of covariance \cite{bCovUrl}, we have for the covariance between $C_\BOne$ and $C_\BTwo$ as
    \begin{equation}
        \label{eCovCb1Cb2Soe}
                \Cov(C_\BOne, C_\BTwo) = B_1 \SumSc(1,|\Delta_b|,\Srb) + B_2 \SumSc(2,|\Delta_b|,\Srb)
    \end{equation}
where $B_1=\Ex^2[\gamma_n] ( A_2^2 + 8 A_2 A_3 \Ex[\gamma_n] + 16 A_3^2 \Ex^2[\gamma_n] )$, $B_2 = 4 A_3^2 \Ex^4[\gamma_n]$, and $\SumSc(r,|\Delta_b|,\Srb)$ is defined in \eqref{eSumSc}.
Thus, we have
    \begin{equation}
        \label{eVarCbSoe}
                \Var[C_b] = B_1 \SumSc(1,0,\Srb) + B_2 \SumSc(2,0,\Srb).
    \end{equation}

\footnote{Personal Note: \Fig{fSumScSumRb}=053, \Fig{fSumScCddSumRbCdd}=054, \Fig{fExMaxCb}=061, \Fig{fExMaxCbCdd}=071, \Fig{fSumRateGainCddvsFreqSel}=073, \Fig{fSumRateGainCddvsSrb}=075, \Fig{fOptFreqSel}=062,074, \Fig{fOptCd}=072}


\bibliographystyle{\TexComDir/IEEEbib}
\bibliography{\TexComDir/MyRefs}

\end{document}